\documentclass[zpreprint,zbstnp,final]{zeus_paper}
\usepackage{placeins}
\usepackage[titletoc,title]{appendix}
\usepackage{authblk}
\usepackage{hyperref}
\usepackage{ifthen}

\newcommand{\SUpsilon}{0.9701\pm0.0284}

\newcommand{\sqrtS}{10.5}
\newcommand{\Rc}{0.4012}
\newcommand{\sigmaeecc}{2399.23}
\newcommand{\Rhadrons}{3.5239}
\newcommand{\RudG}{1.097\pm0.035}
\newcommand{\RudGsign}{2.7}
\newcommand{\EXTRABARYONS}{0.136\pm0.006}
\newcommand{\fitgoodness}{65.6/ 64}
\newcommand{\sigmappcsevenalice}{434\pm 84}
\newcommand{\sigmappctwosevensixalice}{229\pm 67}

\newcommand{\sigmappcsevenatlas}{1400\pm141}

\newcommand{\sigmappcseven}{2689\pm203}
\newcommand{\sigmappcthirteenzero}{5269\pm293}
\newcommand{\sigmappcthirteenone}{4174\pm339}

\newcommand{\Rthirteenseven}{1.96\pm0.18}
\newcommand{\Rseventwosevensix}{1.89\pm0.66}
\newcommand{\SZ}{0.9292\pm0.0261}
\newcommand{\SZsign}{2.7}

\newcommand{\tabEEUmeas}{
\begin{table*}[htbp]\tabfontsize\centering
\begin{tabular}{|c|c|c|c|c|} \hline
\multirow{2}{*}{Decay}   & CLEO        & ARGUS      & BABAR      & BELLE       \\
       & $\sigma \cdot {\cal B},\pb$ & $\sigma,\pb$& $\sigma \cdot {\cal B},\pb$&$\sigma,\pb$  \\
\hline\hline
$D^{*+}\rightarrow D^{0}\pi^+$,    & & & &\\
$    D^{0}\rightarrow  K^-\pi^+$    &$17.0\pm 1.5\pm 1.4$\!\cite{Bortoletto:1988kw}&$690\pm80$\!\cite{Albrecht:1991ss}  &                                             &$598\pm 2\pm 77$\!\cite{Seuster:2005tr} \\
$D^{*+}\rightarrow D^{+}\pi^0$,    & & & &\\
$D^{+}\rightarrow K^-\pi^+\pi^+$ &                                              &                                              &                                             &$590\pm 5\pm 78$\!\cite{Seuster:2005tr} \\
$D^{*+}\rightarrow D^{0}\pi$,    & & & &\\
$D^{0}\rightarrow K^-\pi^-\pi^+\pi^+$   &$33.0\pm 3.0\pm 1.8$\!\cite{Bortoletto:1988kw}&                                              &                                             &                                                \\\hline
$D^{*0}\rightarrow D^0\pi^0/\gamma$          &$30.0\pm 3.0\pm 5.8$\!\cite{Bortoletto:1988kw}&                                              &                                             &                                                \\
$D^{*0}\rightarrow D^0\pi^0$          &                                               &                                              &                                             &$510\pm 3\pm 84$\!\cite{Seuster:2005tr}\\\hline
$D^{0}\rightarrow K^-\pi^+$            &$52 \pm 5 \pm 4$\!\cite{Bortoletto:1988kw}         &$1180\pm150$\!\cite{Albrecht:1991ss}      &                                             &$1449\pm 2\pm 64$\!\cite{Seuster:2005tr}  \\\hline
$D^{+}\rightarrow K^-\pi^+\pi^+$           &$51 \pm 7 \pm 2$\!\cite{Bortoletto:1988kw}       &$650\pm90$\!\cite{Albrecht:1991ss}         &                                             &$654\pm 1\pm 36$\!\cite{Seuster:2005tr} \\

\hline\hline
       & $\sigma \cdot {\cal B},\pb$ & $\sigma \cdot {\cal B},\pb$& $N^{q\bar{q}}_{H} \cdot {\cal B} \times {10^{7}}$&$\sigma,\pb$  \\
\hline\hline

$\Lambda_c^+\rightarrow p K^-\pi^+$         &$13.5\pm 4.0\pm 1.4$\!\cite{Bortoletto:1988kw}&$9.0 \pm 1.2 \pm 1.0$\!\cite{Albrecht:1988an} & $284\pm4\pm9$\!\cite{Aubert:2006cp} &$189\pm 1 \pm 66$\!\cite{Seuster:2005tr}  \\
$\Lambda_c^+\rightarrow p K^-\pi^+$         &$10.0 \pm 1.5 \pm 1.5$\!\cite{Avery:1990bc}   &                                              &                                             &                                                \\
$\Lambda_c^+\rightarrow K^0 p$          &$4.6 \pm 0.6 \pm 0.8$\!\cite{Avery:1990bc}    &                                              &                                             &                                                \\
$\Lambda_c^+\rightarrow K^0 p \pi^+ \pi^-$  &$4.5 \pm 1.3 \pm 1.1$\!\cite{Avery:1990bc}    &                                              &                                             &                                                \\
$\Lambda_c^+\rightarrow \Lambda^0 \pi^+$    &$1.9 \pm 0.3 \pm 0.3$\!\cite{Avery:1990bc}    &                                              &                                             &                                                \\
$\Lambda_c^+\rightarrow \Lambda^0 \pi^+\pi^+\pi^-$   &$6.8 \pm 1.0 \pm 1.3$\!\cite{Avery:1990bc}    &                                              &                                             &                                               \\
$\Lambda_c^+\rightarrow \Xi^- K^+ \pi^+ $     &$1.6 \pm 0.4 \pm 0.3$\!\cite{Avery:1990bc}    &                                              &                                             &                                                \\\hline\hline
       & $\sigma \cdot {\cal B},\pb$ & $\sigma \cdot {\cal B},\pb$& $\sigma \cdot {\cal B},\pb$&$\sigma,\pb$  \\
\hline\hline

$D^+_s\rightarrow \phi\pi^+$            &$7.2\pm 1.9\pm 1.0$\!\cite{Bortoletto:1988kw} &$7.5 \pm 0.8 \pm 0.7$\!\cite{Albrecht:1991pa} &$7.55\pm 0.20\pm 0.34$\!\cite{Aubert:2002ue} &$231\pm 2\pm 92$\!\cite{Seuster:2005tr} \\\hline
$D^{*}_{s}\rightarrow \phi\pi^+\gamma$  &                                                &                                                &$5.8\pm 0.7\pm 0.5$\!\cite{Aubert:2002ue}    &                                                \\\hline

\end{tabular}
\caption{
{Measured  cross-sections, $\sigma$, cross-sections times branching ratios, 
$\sigma \cdot {\cal B}$, and $N^{q\bar{q}}_{H} \cdot {\cal B}$(see text for explanation) 
for the production of the charm hadrons produced in $e^+e^-$ collisions
at centre-of-mass energies of   
$\sqrt{s}=10.55\GeV$\!\protect\cite{Bortoletto:1988kw},
$\sqrt{s}=10.52\GeV$\!\protect\cite{Avery:1990bc,Seuster:2005tr},
$\sqrt{s}=10.47\GeV$\!\protect\cite{Albrecht:1991ss},
$\sqrt{s}=10.5\GeV$\!\protect\cite{Albrecht:1991pa},
$\sqrt{s}=10.2\GeV$\!\protect\cite{Albrecht:1988an} and
$\sqrt{s}=10.54\GeV$\!\protect\cite{Aubert:2006cp,Aubert:2002ue}.
The numbers are given as in the original publications with
uncertainties related to branching ratios omitted.
 The first uncertainty is statistical and the second is systematic. 
}}
\label{tab:EEUmeas}
\end{table*}
}

\newcommand{\tabEEZmeas}{
\begin{table*}[htbp]\tabfontsize\centering
\begin{tabular}{|c|c|c|c|} \hline

\multirow{3}{*}{Particle or decay}   & OPAL & ALEPH & DELPHI \\
           & $\frac{\Gamma_{c\bar c}}{\Gamma_{\text{had}}}\!\cdot\!f(c\rightarrow H)\!\cdot\!{\cal B}$ & $\frac{\Gamma_{c\bar c}}{\Gamma_{\text{had}}}\!\cdot\!f(c\rightarrow H)\!\cdot\!{\cal B}$ & $\frac{\Gamma_{c\bar c}}{\Gamma_{\text{had}}}\!\cdot\!f(c\rightarrow H)\!\cdot\!{\cal B}$ \\
           & ($10^{-6}$) & ($10^{-6}$) & ($10^{-6}$) \\\hline\hline
$D^+_s\rightarrow \phi(K^+K^-)\pi^+$ 
                                   & $560 \pm 150 \pm 70$\!\cite{Alexander:1996wy} 
                                   & $352 \pm 57 \pm 21$\!\cite{Barate:1999bg} 
                                   & $765 \pm 69 \pm 37$\!\cite{Abreu:1999vw} \\
$D^+_s\rightarrow K^{*0}(K\pi)K^+  $ 
								   &                                                            
								   &                                              
								   & $624 \pm 122 \pm 73$\!\cite{Abreu:1999vw} \\\hline

$D^{0}\rightarrow K^-\pi^+  $      
                                   & $3890 \pm 270^{+260}_{-240}$\!\cite{Alexander:1996wy} 
                                   & $3700 \pm 110 \pm 230$\!\cite{Barate:1999bg} 
                                   & $3570 \pm 100 \pm 146$\!\cite{Abreu:1999vw} \\\hline
$D^{+}\rightarrow K^-\pi^+\pi^+  $ 
									& $3580 \pm 460^{+250}_{-310}$\!\cite{Alexander:1996wy} 
									& $3680 \pm 120 \pm 200$\!\cite{Barate:1999bg} 
									& $3494 \pm 116 \pm 140$\!\cite{Abreu:1999vw}\\ \hline
$\Lambda_c^+\rightarrow  p K^- \pi^+$      
									& $410 \pm 190 \pm 70$\!\cite{Alexander:1996wy}
									& $673 \pm 70 \pm 37$\!\cite{Barate:1999bg} 
									& $743 \pm 155 \pm 78$\!\cite{Abreu:1999vw} \\\hline
$D^{*+}\rightarrow D^0(K^-\pi^+)\pi^+$  
									& $1041 \pm 20  \pm 40$\!\cite{Ackerstaff:1997ki}   
									&                                              
									&  $1089\pm 27\pm 39$\!\cite{Abreu:1999vw}\\\hline\hline
                                    & $f(c\rightarrow H)$ & $f(c\rightarrow H)$ & $f(c\rightarrow H)$ \\
                                               & ($10^{-3}$) & ($10^{-3}$) & ($10^{-3}$) \\\hline\hline
$D^{*+} $                           
                                    & $222 \pm14  \pm14$\!\cite{Ackerstaff:1997ki}    
                                    &  $233.3\pm10.2\pm8.41$\!\cite{Barate:1999bg}            
                                    &  \\\hline\hline
                                    & $f(c\rightarrow H) \cdot {\cal B}$ & $f(c\rightarrow H) \cdot {\cal B}$ & $f(c\rightarrow H) \cdot {\cal B}$ \\
                                               & ($10^{-3}$) & ($10^{-3}$) & ($10^{-3}$) \\\hline\hline
$D^{*+} \rightarrow D^0\pi^+$      
                                     &     
                                     & 
                                     & $ 174\pm10\pm 4.2$\!\cite{Abreu:1999vx}  \\\hline
$D^{*+}_s\rightarrow \phi(K^+K^-)\pi^+\gamma$ 
                                     &  
                                     & $69 \pm 18 \pm 7$\!\cite{Barate:1999bg} 
                                     &  \\\hline

\end{tabular}
\caption{LEP measurements of the products of the partial decay width of 
the $Z$ into $c\bar{c}$ quark pairs, $\frac{\Gamma_{c\bar c}}{\Gamma_{\text{had}}}$, charm-hadron-production fractions,
 $f(c \rightarrow H)$, and corresponding branching ratios. 
 The numbers are given as in the original publications with
uncertainties related to branching ratios omitted.
 The first uncertainty is statistical and the second is systematical.
 }
\label{tab:EEZmeas}
\end{table*}
}

\newcommand{\tabDISmeas}{
\begin{table*}[htbp]\tabfontsize\centering
\begin{tabular}{|c|c|c|c|}
\hline
  \multirow{2}{*}{ Decay }                                   
                                           & ZEUS\!\cite{Chekanov:2007ch}  & ZEUS\!\cite{Abramowicz:2010aa}    & H1\!\cite{Aktas:2004ka}   \\
                                           &  $\sigma,\nb$  & $\sigma,\nb$    & $\sigma,\nb$   \\
                                           \hline\hline
$D^{0}\rightarrow K^+\pi^-$             
											& $7.34 \pm 0.36^{+0.35}_{-0.27}\pm 0.13$          
											&                                                                  
											&$6.53\pm 0.49^{+1.06}_{-1.30}$\\
$D^{+}\rightarrow K^+\pi^+\pi^+$             
											& $2.80\pm 0.30^{+0.18}_{-0.14}\pm 0.10$           
											&                                                                  
											&$2.16\pm 0.19^{+0.46}_{-0.35}$\\
$D^{+}\rightarrow K^0_s\pi^+$                                                                                     
											&      
											& $25.7 \pm 4.1{}^{+3.8}_{-5.2}\pm 0.8$
											& \\
$D^+_s\rightarrow \phi(K^+K^-)\pi^+$           
											& $1.27\pm 0.16^{+0.11}_{-0.06}{}^{+0.19}_{-0.15}$
											&                                                                  
											&$1.67\pm 0.41^{+0.54}_{-0.54}$ \\
$D^{*+}\rightarrow D^0(K^-\pi^+)\pi^+$            
											& $3.14\pm 0.12^{+0.18}_{-0.15}\pm 0.06$          
											&                                                                  
											& $2.90\pm 0.20^{+0.58}_{-0.44}$\\
$\Lambda_c^+\rightarrow \Lambda\pi^+$            
											&                                                                       
											&$14.9 \pm 4.9 {}^{+2.2}_{-2.6}\pm 3.9$  
											&  \\
$\Lambda_c^+\rightarrow K^0_sp$            
											&                                                                       
											&$14.0 \pm 5.8 {}^{+3.8}_{-3.3}\pm 3.7$  
											&  \\
$\Lambda_c^+(combination)$
											&                                                                       
											&$14.7 \pm 3.8 {}^{+2.1}_{-2.2}\pm 3.9$  
											&  \\
$D^{0}\rightarrow K^-\pi^+,no\ D^{*+}$   
											& $1.78\pm 0.08^{+0.12}_{-0.10}\pm 0.03$   
											&                                                                  
											&\\\hline
\end{tabular}
\caption{
Measurements of charm-hadron-production cross-sections in DIS in $e^{\pm}p$ collisions.
The numbers are given as in the original publications.
The first uncertainty is statistical, the second is systematical and the third one corresponds to the branching ratio.
}
\label{tab:DISmeas}
\end{table*}
}

\newcommand{\tabPHPmeas}{
\begin{table*}[htbp]\tabfontsize\centering
\begin{tabular}{|c|c|c|}
\hline
    \multirow{2}{*}{ Decay }                       &     ZEUS\!\cite{Chekanov:2005mm}                                                          &  ZEUS\!\cite{Abramowicz:2013eja}         \\
                         &     $\sigma,\nb$                                                           &   $f(c\rightarrow H)$         \\

\hline\hline
$D^{+}\rightarrow K^-\pi^+\pi^+        $ & $5.07\pm0.36{}^{+0.44}_{-0.23}{}^{+0.34}_{-0.30}$      &   $0.234 \pm 0.006{}^{+0.004}_{-0.006}{}^{+0.006}_{-0.008}$\\
$D^{0}\rightarrow K^+\pi^-, no\ D^{*+}  $ & $8.49\pm0.44{}^{+0.47}_{-0.48}{}^{+0.20}_{-0.19}$      &   \\
$D^{0}\rightarrow K^+\pi^-, with\ D^{*+}$ & $2.65\pm0.08{}^{+0.11}_{-0.10}{}\pm0.06$              &     $0.588 \pm 0.017{}^{+0.011}_{-0.006}{}^{+0.012}_{-0.018}$                                                                                  \\
$D^+_s\rightarrow \phi(K^+K^-)\pi^+    $ & $2.37\pm0.20{}\pm 0.20{}^{+0.72}_{-0.45}$             &   $0.088 \pm 0.006{}^{+0.002}_{-0.007}{}^{+0.005}_{-0.005}$ \\
$\Lambda_c^+\rightarrow p K^-\pi^+          $ & $3.59\pm0.66{}^{+0.54}_{-0.66}{}^{+1.15}_{-0.70}$      &   $0.079 \pm 0.013{}^{+0.005}_{-0.009}{}^{+0.024}_{-0.014}$  \\
$D^{*+}\rightarrow D^0(K^-\pi^+)\pi^+     $  &                                                         &   $0.234 \pm 0.006{}^{+0.004}_{-0.004}{}^{+0.005}_{-0.007}$ \\
${}^{\rm add}D^{*+}\rightarrow D^0     $ & $1.05\pm0.07{}^{+0.09}_{-0.04}{}\pm0.03$                   &    \\
${}^{\rm kin}D^{*+}\rightarrow D^0     $ & $4.97\pm0.14{}^{+0.23}_{-0.18}{}^{+0.13}_{-0.12}$          &   \\
\hline
\end{tabular}
\caption{Measurements of charm-hadron-production cross-sections and fragmentation fractions in photoproduction in $e^{\pm}p$ collisions.
 The numbers are given as in the original publications.
The first uncertainty is statistical, the second is systematic and the 
third one corresponds to the branching ratio.}
\label{tab:PHPmeas}
\end{table*}
}

\newcommand{\tabLHCBmeasplusthirteenplusalice}{
\begin{table*}[htpb]\tabfontsize\centering
\begin{tabular}{|c|c|m{0.30em}m{0.30em}m{0.30em}m{0.85em}|c|c|c|c|}\hline
    \multirow{5}{*}{Decay}                          &\multicolumn{5}{c|}{ LHCb\!\cite{Aaij:2013mga} }             & LHCb\!\cite{Aaij:2015bpa}  & ALICE\!\cite{Abelev:2012vra}   &ALICE\!\cite{ALICE:2011aa} & ATLAS\!\cite{Aad:2015zix}  \\
                                                   &    \multicolumn{5}{c|}{$7\TeV$     }       & $13\TeV$               &  $2.76\TeV$                &    $7\TeV$           &  $7\TeV$ \\

                                                   &  \multicolumn{5}{c|}{ $p_T\!\in\![0,8]$ $ y\!\in\![2,4.5]$           }                             &   $p_T\!\in\![1,8]    $    &  $p_T\!\in\![2,12]    $    &  $p_T\!\in\![2,12]    $                  &  $p_T\!\in\![3.5,20]    $    \\                          
                                                   &        &   \multicolumn{4}{c|}{ Corr.(\%)           }                           &   $y\!\in\![2,4.5]$        &  $|y|< 0.5 $    &  $|y|< 0.5     $                  &  $|\eta|< 2.1 $    \\                         
                                                   &  $\sigma,\mb$             &    $D^{0}$&$D^{+}$&$D^{*+}$&$D_s^{+}$                          & $\sigma,\mb$               &  $\sigma,\mb$                &    $\sigma,\mb$                          &  $\sigma,\mb$                 \\
    \hline\hline
    $D^{0}\rightarrow K\pi$                        & $1661\pm  129$            &    &    &    &                                                 &$2460 \pm 3  \pm 130$        & $110 \pm 16 {}^{+28}_{-34}$ &$231 \pm 12 {}_{-56}^{+37}$               &                            \\
    $D^{+}\rightarrow K^-\pi^+\pi^+$               &  $645\pm   74$            & 76 &    &    &                                                 & $1000  \pm 3  \pm 110$      & $47 \pm 9 {}^{+10}_{-12}$  &$81 \pm 7 {}_{-26}^{+23}$                 & $328 \pm 16\pm 27$ \\
    $D^{*+}\rightarrow D^{0}(K^-\pi^+)\pi^+$       &  $677\pm   83$            & 77 & 73 &    &                                                 &$460  \pm 13 \pm 100$        &$59 \pm 14 {}^{+13}_{-14}$  & $104 \pm 6 {}_{-22}^{+17}$               &$331 \pm 18 \pm 28$   \\
    $D^{+}_{s}\rightarrow \phi(K^+K^-)\pi^+$       &  $197\pm   31$            & 55 & 52 & 53 &                                                 & $880  \pm 5  \pm 140$       &                            &    $53 \pm 12{}_{-15}^{+13}$\protect\cite{Abelev:2012tca}             &  $160 \pm 31 \pm 17$  \\
    $\Lambda_c^+\rightarrow p K^-\pi^+$             &  $233\pm   77$           & 26 & 25 & 25 & 18                                              &                             &                           &                                          &                            \\
\hline
                                                   &                           &    &    &    &                                                 &   $p_T\!\in\![0,8]    $     &                             &                                          &                            \\
\hline
    $D^{0}\rightarrow K\pi$                        &                           &    &    &    &                                                 & $3370 \pm 4  \pm 200$       &                              &                                          &                            \\
    $D^{+}\rightarrow K^-\pi^+\pi^+$               &                           &    &    &    &                                                 & $1290 \pm 8  \pm 190$       &                               &                                          &                            \\
    \hline
\end{tabular}
\caption{Measurements of charm-hadron-production cross-sections  in $pp$ collisions.
For the measurements from Ref.~\protect\cite{Aaij:2013mga}  the total uncertainty is given. 
For the rest of measurements the first uncertainty is statistical, the second is systematic. $p_T$ is gived in $\GeV$.}
\label{tab:LHCBmeas}
\end{table*}

}

\newcommand{\tabCSLHCBplusalice}{
\begin{table}[htbp]\tabfontsize\centering

\begin{tabular}{|c|c|c|c|c|}\hline
$\sqrt{s}$,            &  $p_T$               &  $y$ or $\eta$            &     Fit result                 &     Original \\
$\TeV$                 &  range,              &  range                    &     $\sigma(pp\rightarrow c)$, &$\sigma(pp\rightarrow c)$, \\
                       & $\GeV$               &                           &     $\mb$                      &                    $\mb$ \\\hline\hline

${7}$                   &$[0,8]$              &$  y\!\in\![2,4.5]$        &$\sigmappcseven$                &$2838 \pm 268$\!\cite{Aaij:2013mga}\\
${13}$                  &$[1,8]$              &$  y\!\in\![2,4.5]$        &$\sigmappcthirteenone$          &$4300\pm 356$\!\cite{Aaij:2015bpa}\\
${13}$                  &$[0,8]$              &$  y\!\in\![2,4.5]$        &$\sigmappcthirteenzero$         &$5880 \pm 482$\!\cite{Aaij:2015bpa}\\\hline\hline
${2.76}$                &$[2,12]$             &$  |y|< 0.5$             &$\sigmappctwosevensixalice$     &\\
${7}$                   &$[2,12]$             &$  |y|< 0.5$             &$\sigmappcsevenalice$           &\\\hline\hline
${7}$                   &$[3.5,20]$           &$|\eta|< 2.1$            &$\sigmappcsevenatlas$           &\\
\hline
\end{tabular}
\caption{
The values of the  inclusive fiducial charm quark production cross-section, $\sigma(pp\rightarrow c)$, from the original publications and obtained in the global fit.
The statistical,
systematic and fragmentation uncertainties of the values from Refs.~\protect\cite{Aaij:2013mga,Aaij:2015bpa}  were added in quadrature.
}
\label{tab:CSLHCB}
\end{table}
}

\newcommand{\tabexcited}{
\begin{table*}[htbp]\tabfontsize\centering
\begin{tabular}{|c|c|c|c|}\hline
   \multirow{3}{*}{ Particle }                            &                  ZEUS          & OPAL                            & ALEPH \\
                             &   $f(c\rightarrow H)$          & $f(c\rightarrow H)$             &$f(c\rightarrow H)$ \\
                             &   ($10^{-2}$)                  & ($10^{-2}$)                     &($10^{-2}$)              \\\hline\hline
$D^0_{1}$    &   $2.9\pm0.5{}^{+0.5}_{-0.5}$\!\cite{Abramowicz:2012ys}  & $2.1\pm0.7\pm0.3$\!\cite{Ackerstaff:1997vc}     & \\
                             &   $3.5\pm0.4^{+0.4}_{-0.6}$\!\cite{Chekanov:2008ac}    &      & \\\hline
$D^{*0}_{2}$ &   $3.9\pm0.9{}^{+0.8}_{-0.6}$\!\cite{Abramowicz:2012ys}     & $5.2\pm2.2\pm1.3$\!\cite{Ackerstaff:1997vc}     &\\
                             &   $3.8\pm0.7^{+0.5}_{-0.6}$\!\cite{Chekanov:2008ac}  &                                                 &\\\hline
$D^+_{1}$    &   $4.6\pm1.8{}^{+2.0}_{-0.3}$\!\cite{Abramowicz:2012ys}     &                       &\\\hline
$D^{*+}_{2}$ &   $3.2\pm0.8{}^{+0.5}_{-0.2}$\!\cite{Abramowicz:2012ys}   &                       &\\\hline
$D^+_{s1}$   &   $1.11\pm0.16^{+0.08}_{-0.10}$\!\cite{Chekanov:2008ac}   & $1.6\pm0.4\pm0.3$\!\cite{Ackerstaff:1997vc}     & $0.94\pm0.22\pm0.07$\!\cite{Heister:2001nj} \\\hline
\end{tabular}
\caption[Comparison of $D^{**}$ fragmentation-fractions results]{
Comparison of fragmentation-fraction-results of  measurements of excited charm mesons.}
\label{tab:excited}
\end{table*} 
}

\newcommand{\figFINALaverageFF} {
\begin{figure}[!htb]
\begin{center}
\includegraphics[width=0.99\linewidth]{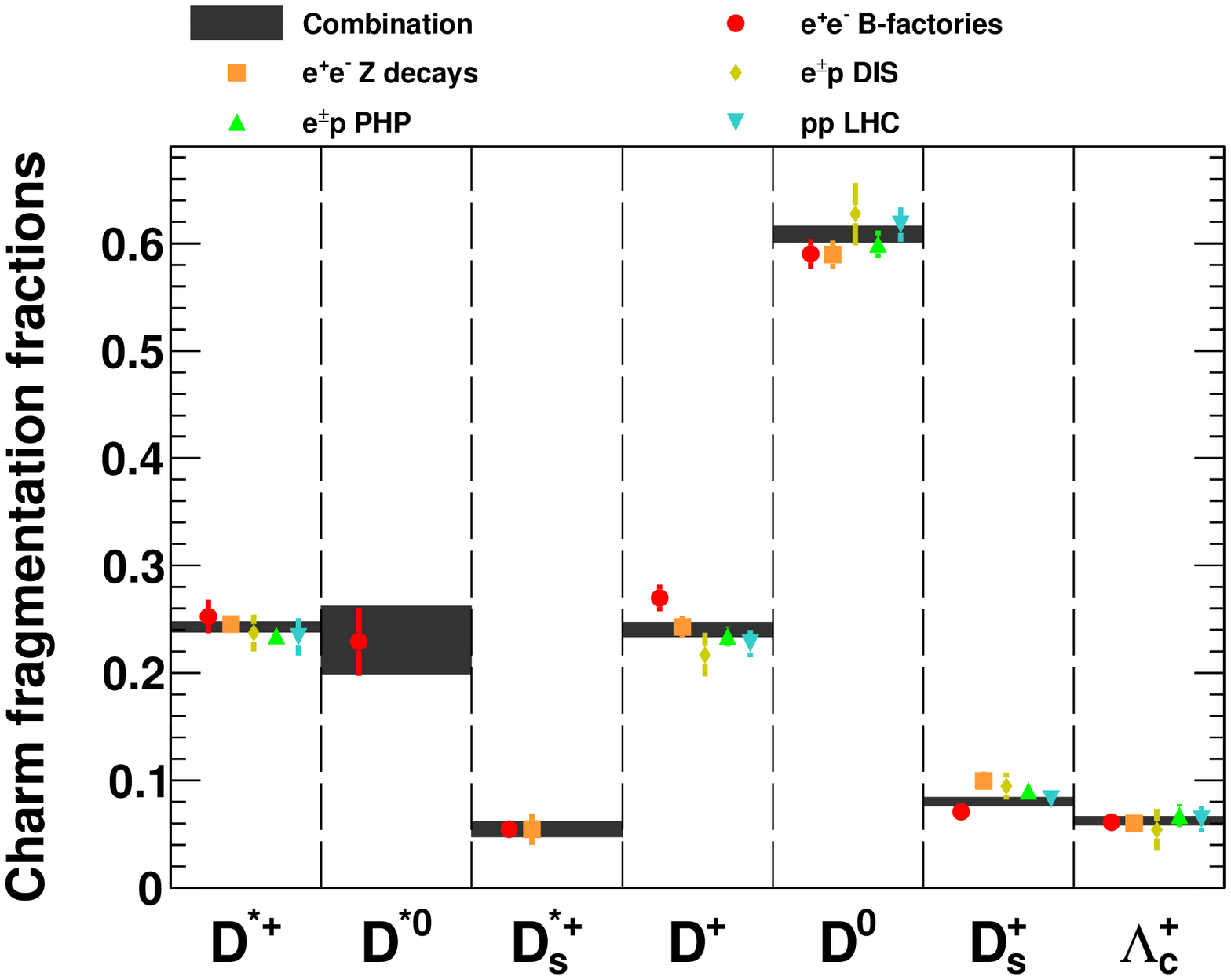}
\caption{The values of charm-quark fragmentation fractions, $f(c\rightarrow H)$,
in different experiments with the $S$ constraint. 
The global combination with the $S$ constraint is shown with the shaded  band. 
Averages of included data in different production regimes are shown with various full symbols. }
\label{fig:FINALaverageFF}
\end{center}
\end{figure}
}
\newcommand{\figFINALaverageRPg} {
\begin{figure}[!htb]
\begin{center}
\includegraphics[width=0.99\linewidth]{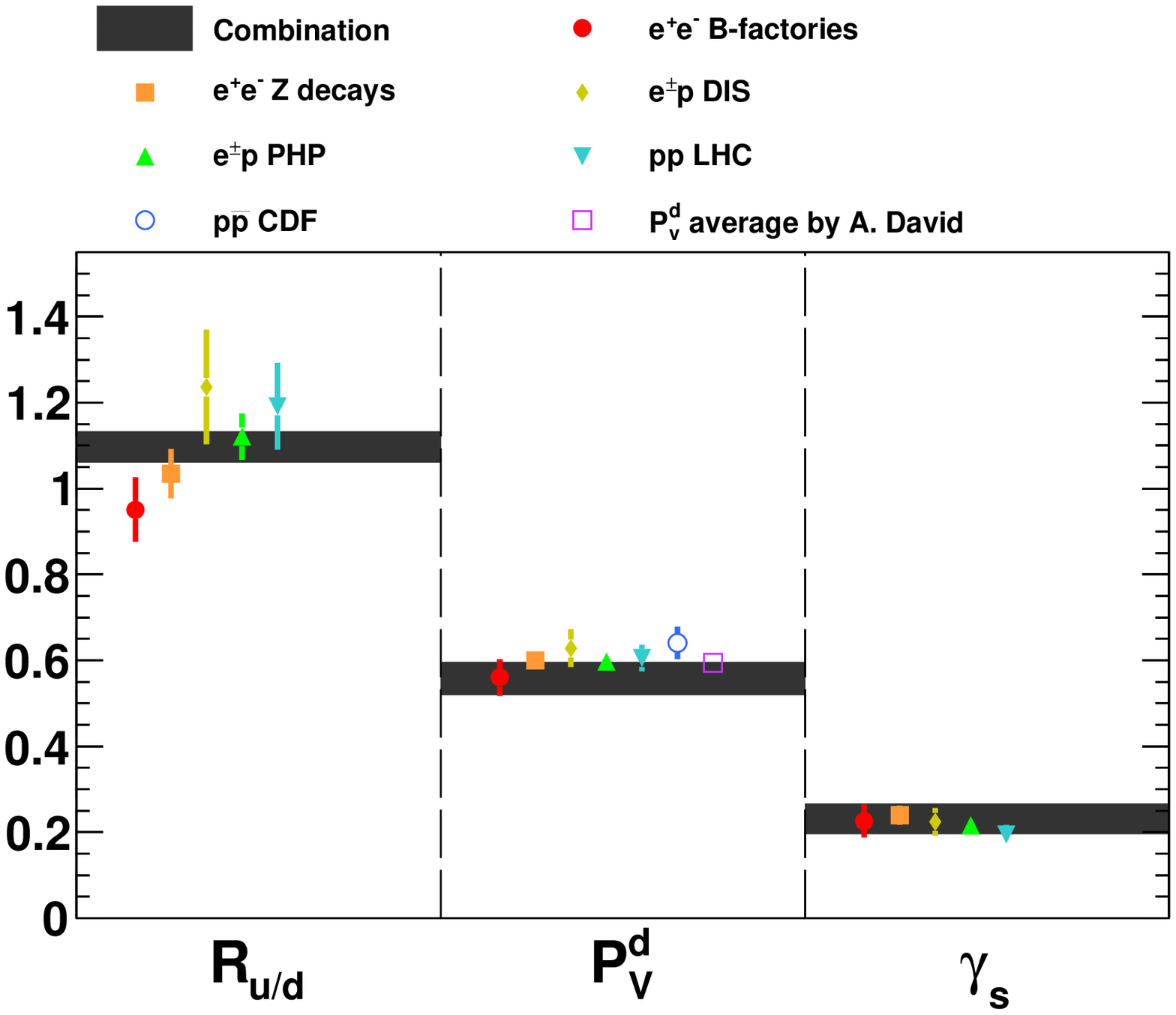}
\caption{The values of $R_{u/d}$, $P^d_V$ and  $\gamma_s$
in different experiments with the $S$ constraint. 
The global combination with the S constraint is shown with the shaded  band. 
Combinations of included data in different production regimes are shown with various full symbols. 
Data that were not included in the combination~\protect\cite{
Acosta:2003ax, David:2007iv} 
are shown with open symbols. 
Note, that the latter are quoted from the original papers, 
i.e. without correction to the up-to-date branching ratios 
and with no branching ratio uncertainty, if not given in the source.}
\label{fig:FINALaverageRPg}
\end{center}
\end{figure}
}

\title{Combined analysis of charm-quark fragmentation-fraction measurements}
\author[a]{Mykhailo Lisovyi\thanks{mikhaylo.lisovyi@desy.de}}
\author[b]{Andrii Verbytskyi\thanks{andrii.verbytskyi@mpp.mpg.de}}
\author[c]{Oleksandr Zenaiev\thanks{oleksandr.zenaiev@desy.de}}

\affil[a]{Physikalisches Institut der Universit\"{a}t Heidelberg
}
\affil[b]{Max-Planck-Institut f\"{u}r Physik
}
\affil[c]{on leave from DESY
}
\prepnum{MPP-2015-257}

\abstract{
A summary of measurements of the fragmentation of charm quarks into a
specific hadron is given. Measurements performed in photoproduction and deep
inelastic scattering in $e^{\pm}p$, $pp$ and $e^+e^-$ collisions
are compared, using up-to-date branching ratios. Within uncertainties, 
all measurements agree, supporting the hypothesis that fragmentation 
is independent of the specific production process. Averages of the 
fragmentation fractions over all measurements are presented.
The average has significantly reduced uncertainties
compared to individual measurements.

}

\newcommand{\eVdist}{\kern-0.06667em}
\newcommand{\GeV}{{\,\text{Ge}\eVdist\text{V\/}}}
\newcommand{\TeV}{{\,\text{Te}\eVdist\text{V\/}}}

\newcommand{\MeV}{{\,\text{Me}\eVdist\text{V\/}}}
\newcommand{\nb}{{\,\text{nb}}}
\newcommand{\mb}{{\,\mu \text{b}}}
\newcommand{\pb}{{\,\text{pb}}}
\newcommand{\epem}{$e^+e^-\,$}
\newcommand{\ep}{$e^{\pm}p\,$}
\newcommand{\fig}[1]{Fig.~\ref{fig:#1}}
\newcommand{\tab}[1]{Tab.~\ref{tab:#1}}
\newcommand{\sect}[1]{Sec.~\ref{sec:#1}}
\newcommand{\app}[1]{App.~\ref{sec:#1}}
\newcommand{\eq}[1]{Eq.~(\ref{eq:#1})}
\newcommand{\FB}{\FloatBarrier}
\newcommand{\epjcbreak}[1]{}
\newcommand{\draftbreak}[1]{}
\newcommand{\arxivbreak}[1]{\\#1}
\newcommand{\tabfontsize}{\scriptsize}

\begin{document}
\makezeustitle\newpage
\clearpage
\pagenumbering{arabic}
\pagestyle{plain}
\newpage
%%%%%%%%%%%%%%%%%%%%%%%%%%%%%%%%%%%%%%%%%%%%%%%%%%%%%%%%%%%%%%%%%%%%%%%%
\section{Introduction}
\label{sec:int}
The production of specific charm hadrons has
been measured in different regimes and environments: in 
$e^+e^-$  collisions at $B$-factories~\cite{Bortoletto:1988kw,
Avery:1990bc,Albrecht:1991ss,Albrecht:1991pa,
Albrecht:1988an,Aubert:2002ue,
Seuster:2005tr,Aubert:2006cp}  and in $Z$
decays~\cite{Alexander:1996wy,Ackerstaff:1997ki,Barate:1999bg, 
Abreu:1999vw,Abreu:1999vx}, in \ep collisions in photoproduction 
(PHP)~\cite{Chekanov:2005mm,Abramowicz:2013eja},  deep inelastic 
scattering (DIS)~\cite{Chekanov:2007ch,
Abramowicz:2010aa,Aktas:2004ka} and in
$pp$ collisions~\cite{Aaij:2013mga, Aaij:2015bpa,ALICE:2011aa,
Abelev:2012tca,Abelev:2012vra,Aad:2015zix}.

The fragmentation process is soft and hence can not be calculated with
the techniques of perturbative QCD (pQCD).
Therefore, these measurements 
are a necessary ingredient for any QCD prediction of 
charm-hadron production.
In this context, it is important to validate the hypothesis 
that fragmentation 
fractions are universal, i.e.\ independent of the hard production 
mechanism. Thus, once precisely measured in one experiment, they can 
be applied in any reaction.
Another important check is that the sum of fragmentation fractions of 
all known weakly decaying charm hadrons is equal to unity, thus checking
 if all weakly decaying states are known.

To achieve these goals, a comparison of fragmentation-fraction 
measurements obtained in different production regimes 
is performed using a combination of individual measurements.
Due to independent data sets and different detector types and 
constructions, the experimental statistical and systematic uncertainties 
in most cases can be treated as uncorrelated between measurements.
However, a careful treatment of correlated uncertainties due to common 
usage of
branching ratios and theory inputs is essential, as in many measurements 
these are one of the leading uncertainty sources.
In the past several combinations of fragmentation-fraction data were 
performed with fewer inputs: the summary of the charm fragmentation 
fractions in $e^+e^-$ at the $Z$ resonance~\cite{PDG2014}, the combination 
of $e^+e^-$ measurements~\cite{Gladilin:1999pj,Gladilin:2014tba} as well as 
the combination of $e^+e^-$ and $e^{\pm}p$ 
measurements~\cite{Lohrmann:2011np}.
Compared to those, the present analysis extends to a larger set of 
measurements, in particular the final measurement in PHP by the ZEUS 
experiment at HERA~\cite{Abramowicz:2013eja},  the $pp$ measurements 
from LHCb~\cite{Aaij:2013mga, Aaij:2015bpa}, 
ALICE~\cite{ALICE:2011aa,Abelev:2012tca,Abelev:2012vra}, ATLAS~\cite{Aad:2015zix}
 and the $\Lambda_c^+$ measurements from 
the BABAR experiment~\cite{Aubert:2006cp}. It uses the up-to-date 
branching-ratio values~\cite{PDG2014,Ablikim:2014mww,
Aubert:2005ik}, treats correlations of 
branching-ratio uncertainties and recent theory predictions with 
reduced uncertainties~\cite{Chetyrkin:2000zk,Freitas:2014hra} as input.

%%%%%%%%%%%%%%%%%%%%%%%%%%%%%%%%%%%%%%%%%%%%%%%%%%%%%%%%%%%%%%%%%%%%%%%%
\section{Combination of individual measurements}
\label{sec:comb}
\subsection{Update of input measurements to recent branching ratios}
\label{sec:pdgupdate}
To make separate inputs consistent, the original measurements are 
corrected to the same up-to-date world averages of branching ratios  of 
the charm-hadron decays, ${\cal B}$, 
summarised in Tab.~\ref{tab:BIGPDG}. Most of 
the values were taken from Ref.~\cite{PDG2014}.
The ${\cal B}(D^{*0} \rightarrow D^{0}\pi^0)$ was calculated from the 
two most precise measurements~\cite{Ablikim:2014mww,Aubert:2005ik} of 
${\cal B}(D^{*0} \rightarrow D^{0}\pi^0)/{\cal B}(D^{*0} 
\rightarrow D^{0}\gamma)$ assuming ${\cal B}(D^{*0} \rightarrow 
D^{0}\pi^0)+{\cal B}(D^{*0} $\epjcbreak{}$\rightarrow D^{0}\gamma)=1$.

\begin{table*}[htbp]\tabfontsize\centering
\begin{tabular}{|c|c|c|}\hline
Decay      &  In this work (\%) & In experiments (\%)\\\hline\hline
$D^{*+}\rightarrow{}D^+             $&$32.30\pm0.50$~\cite{PDG2014}                          &                                                                                                               \\
\hline
$D^{*+}\rightarrow{}D^+\pi^{0}      $&$30.70\pm0.50$~\cite{PDG2014}                          &$30.70\pm0.50$,BELLE~\cite{Seuster:2005tr}                                                                     \\
\hline
$D^{*+}\rightarrow{}D^0\pi^{+}      $&$67.70\pm0.50$~\cite{PDG2014}                          &$68.13\pm1.40$,ALEPH~\cite{Barate:1999bg}               $67.70\pm0.50$,ALICE~\cite{ALICE:2011aa,Abelev:2012tca,Abelev:2012vra}\\
$                                   $&                                                       &$55.00\pm4.00$,ARGUS~\cite{Albrecht:1988an,Albrecht:1991pa,Albrecht:1991ss} $67.70\pm0.50$,ATLAS~\cite{Aad:2015zix}                \\
$                                   $&                                                       &$67.70\pm0.50$,BELLE~\cite{Seuster:2005tr}              $67.60\pm0.50$,H1~\cite{Aktas:2004ka}                  \\
$                                   $&                                                       &$67.70\pm0.50$,ZEUS~\cite{Chekanov:2007ch,Chekanov:2008ac,Chekanov:2005mm,Abramowicz:2013eja} $67.70\pm0.50$,LHCb~\cite{Aaij:2015bpa}                \\
$                                   $&                                                       &$68.30\pm1.40$,OPAL~\cite{Ackerstaff:1997ki,Alexander:1996wy}                                                        \\
\hline
$D^{*0}\rightarrow{}D^0             $&$100.00$~\cite{Ablikim:2014mww,Aubert:2005ik}          &$100.00$,H1~\cite{Aktas:2004ka}                         $100.00$,ZEUS~\cite{Chekanov:2007ch}                   \\
\hline
$D^{*0}\rightarrow{}D^0\pi^{0}      $&$64.94\pm0.89$~\cite{Ablikim:2014mww,Aubert:2005ik}    &$61.90\pm2.90$,BELLE~\cite{Seuster:2005tr}                                                                     \\
\hline
$D^{*+}_{s}\rightarrow{}D^{+}_{s}\gamma$&$93.50\pm0.70$~\cite{PDG2014}                          &$94.20\pm0.70$,BABAR~\cite{Aubert:2002ue,Aubert:2006cp}                                                        \\
\hline
$D^+\rightarrow{}K^{-}\pi^{+}\pi^{+}$&$9.46\pm0.24$~\cite{PDG2014}                           &$9.13\pm0.19$,ALICE~\cite{ALICE:2011aa,Abelev:2012tca,Abelev:2012vra} $7.70\pm1.00$,ARGUS~\cite{Albrecht:1988an,Albrecht:1991pa,Albrecht:1991ss}\\
$                                   $&                                                       &$9.13\pm0.19$,ATLAS~\cite{Aad:2015zix}                  $9.20\pm0.60$,BELLE~\cite{Seuster:2005tr}              \\
$                                   $&                                                       &$9.00\pm0.60$,H1~\cite{Aktas:2004ka}                    $9.51\pm0.34$,ZEUS~\cite{Chekanov:2007ch,Chekanov:2008ac}\\
$                                   $&                                                       &$9.20\pm0.60$,ZEUS~\cite{Chekanov:2005mm}               $9.13\pm0.19$,ZEUS~\cite{Abramowicz:2013eja}           \\
$                                   $&                                                       &$9.13\pm0.19$,LHCb~\cite{Aaij:2015bpa,Aaij:2013mga}                                                            \\
\hline
$D^+\rightarrow{}K^{0}\pi^{+}       $&$1.53\pm0.06$~\cite{PDG2014}                           &                                                                                                               \\
\hline
$D^0\rightarrow{}K^{-}\pi^{+}       $&$3.93\pm0.04$~\cite{PDG2014}                           &$3.85\pm0.09$,ALEPH~\cite{Barate:1999bg}                $3.87\pm0.05$,ALICE~\cite{ALICE:2011aa,Abelev:2012tca,Abelev:2012vra}\\
$                                   $&                                                       &$3.71\pm0.25$,ARGUS~\cite{Albrecht:1988an,Albrecht:1991pa,Albrecht:1991ss} $3.88\pm0.05$,ATLAS~\cite{Aad:2015zix}                 \\
$                                   $&                                                       &$3.80\pm0.09$,BELLE~\cite{Seuster:2005tr}               $3.83\pm0.09$,H1~\cite{Aktas:2004ka}                   \\
$                                   $&                                                       &$3.80\pm0.07$,ZEUS~\cite{Chekanov:2007ch,Chekanov:2008ac} $3.80\pm0.90$,ZEUS~\cite{Chekanov:2005mm}              \\
$                                   $&                                                       &$3.88\pm0.05$,ZEUS~\cite{Abramowicz:2013eja}            $3.89\pm0.05$,LHCb~\cite{Aaij:2015bpa,Aaij:2013mga}    \\
\hline
$D^0\rightarrow{}K^{-}\pi^{+}\pi^{+}\pi^{-}$&$8.07\pm0.23$~\cite{PDG2014}                           &                                                                                                               \\
\hline
$D^{+}_{s}\rightarrow{}K^{*0}(K^{-}\pi^{+})K^{+}$&$2.61\pm0.09$~\cite{PDG2014}                           &                                                                                                               \\
\hline
$D^{+}_{s}\rightarrow{}\phi(K^{+}K^{-})\pi^{+}$&$2.27\pm0.08$~\cite{PDG2014}                           &$2.28\pm0.12$,ALICE~\cite{ALICE:2011aa,Abelev:2012tca}  $2.16\pm0.28$,ZEUS~\cite{Chekanov:2007ch}              \\
$                                   $&                                                       &$2.28\pm0.12$,ZEUS~\cite{Abramowicz:2013eja}            $0.00$,LHCb~\cite{Aaij:2015bpa,Aaij:2013mga}           \\
\hline
$D^{+}_{s}\rightarrow{}\phi\pi^{+}  $&$4.50\pm0.40$~\cite{PDG2014}                           &$3.60\pm0.90$,ALEPH~\cite{Barate:1999bg}                $3.60\pm0.90$,BELLE~\cite{Seuster:2005tr}              \\
$                                   $&                                                       &$3.60\pm0.90$,H1~\cite{Aktas:2004ka}                    $3.60\pm0.90$,ZEUS~\cite{Chekanov:2005mm}              \\
\hline
$D^{+}_{s}\rightarrow{}K^{+}K^{-}\pi^{+},7\MeV$&$1.83\pm0.06$~\cite{PDG2014,delAmoSanchez:2010yp}      &$1.85\pm0.11$,ATLAS~\cite{Aad:2015zix}                                                                         \\
\hline
$D^{+}_{s}\rightarrow{}K^{+}K^{-}\pi^{+},20\MeV$&$2.23\pm0.08$~\cite{PDG2014,delAmoSanchez:2010yp}      &$2.24\pm0.13$,LHCb~\cite{Aaij:2015bpa,Aaij:2013mga}                                                            \\
\hline
$K^{0}\rightarrow{}\pi^{+}\pi^{-}   $&$69.20\pm0.05$~\cite{PDG2014}                          &                                                                                                               \\
\hline
$K^{*0}\rightarrow{}K^{+}\pi^{-}    $&$66.67$~\cite{PDG2014}                                 &$66.67$,DELPHI~\cite{Barate:1999bg}                                                                            \\
\hline
$\phi\rightarrow{}K^{+}K^{-}        $&$48.90\pm0.50$~\cite{PDG2014}                          &$49.10\pm0.80$,ALEPH~\cite{Barate:1999bg}               $49.00\pm1.00$,ARGUS~\cite{Albrecht:1988an,Albrecht:1991pa,Albrecht:1991ss}\\
$                                   $&                                                       &$48.90\pm0.50$,BABAR~\cite{Aubert:2002ue,Aubert:2006cp} $49.10\pm0.60$,BELLE~\cite{Seuster:2005tr}             \\
$                                   $&                                                       &$49.00\pm1.00$,CLEO~\cite{Avery:1990bc,Bortoletto:1988kw} $49.10\pm0.80$,DELPHI~\cite{Barate:1999bg}             \\
$                                   $&                                                       &$49.20\pm0.70$,H1~\cite{Aktas:2004ka}                   $49.10\pm0.60$,ZEUS~\cite{Chekanov:2005mm}             \\
$                                   $&                                                       &$49.10\pm0.80$,OPAL~\cite{Ackerstaff:1997ki,Alexander:1996wy}                                                        \\
\hline
$\Lambda^{0}\rightarrow{}p\pi^{+}   $&$63.90\pm0.50$~\cite{PDG2014}                          &                                                                                                               \\
\hline
$\Lambda_c^{+}\rightarrow{}pK^{-}\pi^{+}$&$6.84{}^{+0.32}_{-0.40}$~\cite{PDG2014}                &$5.00\pm1.30$,BABAR~\cite{Aubert:2002ue,Aubert:2006cp}  $5.00\pm1.30$,BELLE~\cite{Seuster:2005tr}              \\
$                                   $&                                                       &$5.00\pm1.30$,ZEUS~\cite{Abramowicz:2010aa,Chekanov:2005mm,Abramowicz:2013eja} $5.00\pm1.30$,LHCb~\cite{Aaij:2013mga}                 \\
\hline
$\Lambda_c^{+}\rightarrow{}\Lambda^{0}(p\pi^{+})\pi^{+}$&$0.95\pm0.10\pm0.05$~\cite{PDG2014}                    &$0.68\pm0.18$,ZEUS~\cite{Abramowicz:2010aa}                                                                    \\
\hline
$\Lambda_c^{+}\rightarrow{}\Lambda^{0}\pi^{+}\pi^{+}\pi^{-}$&$3.59\pm0.28$~\cite{PDG2014}                           &                                                                                                               \\
\hline
$\Lambda_c^{+}\rightarrow{}\Lambda^{0}\pi^{+}$&$1.46\pm0.13$~\cite{PDG2014}                           &                                                                                                               \\
\hline
$\Lambda_c^{+}\rightarrow{}\Xi^{-}K^{+}\pi^{+}$&$0.70\pm0.08$~\cite{PDG2014}                           &                                                                                                               \\
\hline
$\Lambda_c^{+}\rightarrow{}pK^{0}   $&$3.21\pm0.30$~\cite{PDG2014}                           &                                                                                                               \\
\hline
$\Lambda_c^{+}\rightarrow{}pK^{0}(\pi^{+}\pi^{-})$&$1.09\pm0.06\pm0.11$~\cite{PDG2014}                    &$0.80\pm0.21$,ZEUS~\cite{Abramowicz:2010aa}                                                                    \\
\hline
$\Lambda_c^{+}\rightarrow{}pK^{0}\pi^{+}\pi^{-}$&$3.50\pm0.40$~\cite{PDG2014}                           &                                                                                                               \\
\hline

\end{tabular}
\caption{
Branching ratios  used for calculations. 
The second uncertainty for 
the ${\cal B}(\Lambda_c^+)$ is the uncertainty 
of decay branching ratios of daughters.
% A small scalar contribution~\protect\cite{delAmoSanchez:2010yp}  about 5\% in $D^+_s\rightarrow\pi^{+}\phi(K^+K^-)$ decay 
%is neglected.
The numbers in the $D^+_s$ decay branching ratio indicate the used $|M(K^+K^+)-M(\phi(1020))|$ mass windows.
For the experiments which measured combination of cross-sections and branching ratios,
the values of branching ratios are not given.
}
\label{tab:BIGPDG}
\end{table*}

\subsection{Calculation of the fragmentation fractions}
\label{sec:comb::ffdef}
In this paper the charm-quark  fragmentation fraction to a specific 
hadron $H$ is defined as the production cross-section via  charm 
quark divided by the production cross-section of the charm quark:
%-----------------------------------------------------------------------
\begin{equation}\label{eq:ffwithcharmcs}
f(c \rightarrow H) = {\sigma(H)}/{\sigma(c)}. 
\end{equation}
%-----------------------------------------------------------------------
The charm hadrons produced in the decays of beauty hadrons are not 
considered.
The Standard Model makes precise predictions for the total charm 
cross-section in $e^+e^-$ collisions, therefore, for those processes it 
is possible to calculate $f(c \rightarrow H)$ according to 
\eq{ffwithcharmcs}.
Sufficiently precise predictions for the charm-quark production 
in $pp$ and $e^{\pm}p$ collisions are not available.
However, it is possible to make an assumption that the sum of charm-quark 
fragmentation  fractions to all known weakly decaying charm hadrons
in the end of the fragmentation process is unity. 
Then the 
charm-quark fragmentation fraction to a specific hadron can be 
calculated as the ratio of the hadron-production cross-section over the 
sum of cross-sections of all known weakly decaying (w.d.)  charm hadrons
%-----------------------------------------------------------------------
\begin{equation}\label{eq:ffwithsumh}
f(c\rightarrow H)={\sigma(H)}/{\sum_{w.d.}\sigma(H)}. 
\end{equation}
%-----------------------------------------------------------------------

To obtain the charm-quark fragmentation fractions according to 
\eq{ffwithsumh}, in addition to the production cross-sections of $D$ 
mesons and $\Lambda_c^+$, it is necessary to know the cross-sections 
of the weakly decaying $\Xi_c^{+,0}$ and $\Omega_c^0$ states.
Those states are poorly studied, therefore as in 
Ref.~\cite{Alexander:1996wy} it is assumed that ratios of fragmentation 
fractions of charm and strange quarks into the corresponding baryons are 
similar,
 $f(c\rightarrow \Xi_c^+)/f(c\rightarrow \Lambda_c^+)=
  f(c\rightarrow \Xi_c^0)/f(c\rightarrow \Lambda_c^+)=
  f(s\rightarrow \Xi^-)/f(s\rightarrow \Lambda^0)$
and  
 $f(c\rightarrow \Omega_c^0)/f(c\rightarrow \Lambda_c^+)=
  f(s\rightarrow \Omega^-)/f(s\rightarrow \Lambda^0).
  $
In this approach the sum of the production cross-sections of these 
states can be estimated as
\begin{equation}
\sigma(\Xi_c^+)+\sigma(\Xi_c^0)+\sigma(\Omega_c^0)
=\lambda\sigma(\Lambda_c^+),
\end{equation}
where we define
\begin{equation}
\lambda=2\frac{f(s\rightarrow \Xi^-)}{f(s\rightarrow \Lambda^0)}+
\frac{f(s\rightarrow \Omega^-)}{f(s\rightarrow\Lambda^0)}=\EXTRABARYONS.
\end{equation}
%-----------------------------------------------------------------------
The  value of $\lambda$ is 
calculated using the most precise set of $s$ quark fragmentation 
fractions
$f(s\rightarrow \Xi^-)=0.0016\pm 0.0003$,
$f(s\rightarrow \Omega^-)=0.0258\pm 0.0010$ and 
$f(s\rightarrow \Lambda^0)=0.3915\pm 0.0065$ from Ref.~\cite{PDG2014} 
obtained at LEP.
Hereby, the sum of production cross-sections of all weakly decaying 
states is
%-----------------------------------------------------------------------
\begin{multline}
\label{eq:lambda}
\sum_{w.d.}\sigma(H)=\sigma(D^0)+\sigma(D^+)+\sigma(D_s^+)+
\sigma(\Lambda_c^+)+\epjcbreak{+}\lambda \sigma(\Lambda_c^+).
\end{multline}

The fragmentation fractions calculated according to \eq{ffwithcharmcs} 
for the $e^+e^-$ collisions and $Z$ decays 
allow an independent check that
%-----------------------------------------------------------------------
\begin{multline}\label{eq:ffsumisone}
S=f(c\rightarrow D^0)+f(c\rightarrow D^+)+f(c\rightarrow D_s^+)+
f(c\rightarrow \Lambda_c^+)+\epjcbreak{+}\lambda f(c\rightarrow 
\Lambda_c^+)
\end{multline}
%-----------------------------------------------------------------------
is close to unity with sufficient accuracy.
\subsection{Combination procedure}
\label{sec:comb:proc}
The combination of the measurements used in the present analysis is 
based on numerical $\chi^2$ minimisation with respect to observables of 
interest.
The numerical minimisation was performed with the MINUIT 
package~\cite{minuit} and the procedure for calculation of $\chi^2$ 
itself is outlined below.

For a set of $m$ measurements and corresponding expectation values
 calculated 
from $n$ parameters, a column-vector of the residuals $R(1\times m)$ is 
calculated as a difference of a measurement and the corresponding 
expectation.  The covariance matrix, $V(m \times m)$,  is calculated  as 
%-----------------------------------------------------------------------
$$V_{ij}=U^2_{i}\delta_{ij}+\sum_{k}C_{j,k}C_{i,k},$$
%-----------------------------------------------------------------------
where $U_{i}$ stands for an uncorrelated uncertainty of $i$-th residual, 
$C_{i,k}$ stands for the correlated uncertainty of source $k$ of the 
$i$-th measurement and the sum runs over all sources of correlated 
uncertainties.
The $\chi^2$ is then calculated as 
%-----------------------------------------------------------------------
$$\chi^2=R^{T}V^{-1}R.$$
%-----------------------------------------------------------------------

The correlated uncertainties
are treated multiplicatively in the
construction of the covariance matrix, i.e.\ the relative uncertainties 
are used to scale the corresponding expectation values 
instead of the measurement.
This avoids the bias for normalisation uncertainties, such as branching 
ratio uncertainties, which are the main correlated uncertainties 
considered in the presented analysis.
The statistical and uncorrelated systematic uncertainties are treated 
additively.
Data sets and their systematic uncertainties are  assumed to be 
independent between experiments.
In addition, most of the measurements do not contain the information 
about a potential correlation between cross-section values for different 
charm hadrons.
Therefore, in the following all experimental uncertainties  are treated 
as uncorrelated, unless otherwise stated. Uncertainties on the combined 
values of the fragmentation fractions are determined using the Hessian 
method with the criterion $\Delta\chi^2=1$\footnote{
As an illustrative example, when
for a given combination set-up the inputs are the
$m$ cross-section measurements. 
These data define the $m \times m$ covariance matrix $V$ with the 
experimental statistical and systematic uncertainties contributing 
to the diagonal elements 
and the correlated uncertainties setting the off-diagonal elements 
and contributing to the total uncertainties on the diagonal.
The correlated uncertainties considered are those related to
$\lambda$ in \eq{lambda} and branching ratios.
The residuals are obtained subtracting from the measurements 
cross-section expectations calculated from $n$ free parameters
in the fit, which could be fragmentation fractions, 
total charm cross-sections, kinematic factors, etc.
The details of this calculation are outlined in the each section.
With all these components at hand the $\chi^2$ can be evaluated 
and iteratively numerically minimised with respect to the 
free parameters.}.

The evaluated total uncertainties on the free parameters comprise 
experimental, branching ratio and  uncertainties of the $\lambda$ 
parameter.

The combination of all the measurements  is obtained imposing the 
normalisation constraint on the sum of all ground state hadrons by 
adding an additional ``measurement'' of $S$ calculated from 
\eq{ffsumisone} with an uncertainty on $\lambda$ and the  corresponding 
prediction $S = 1$.
In order to keep the main result with the normalisation constraint as 
model independent as possible, no theory inputs on the charm 
cross-section are used in  such  a combination, and the fragmentation 
fractions are calculated according to
  \eq{lambda}.
For the same reason, any measurements that require theoretical inputs 
for conversion into cross-sections or fragmentation fractions and do not 
have these inputs in the original publications are also excluded from 
the main combination.
However, such data are included in a more constrained 
combination.
In the following,  treatment of such measurements will be discussed 
case-by-case in the relevant sections.

The quantities commonly used as Monte Carlo generator parameters,
%-----------------------------------------------------------------------
$$
R_{u/d} =\frac{f(c\rightarrow c\bar{u})}{f(c\rightarrow c\bar{d})}
\approx\frac{f(c\rightarrow D^{0})
-f(c\rightarrow D^{*+}){\cal B}_{D^{*+}\rightarrow D^0} 
}
{
f(c\rightarrow D^{+})
+f(c\rightarrow D^{*+}){\cal B}_{D^{*+}\rightarrow D^0} 
},
$$
$$
\gamma_{s}=\frac{2f(c\rightarrow c\bar{s})}{f(c\rightarrow c\bar{u}/
\bar{d})}(J=0)\approx\frac{2f(c\rightarrow D^+_{s})}{f(c\rightarrow D^+)
+f(c\rightarrow D^0)},
$$ 
$$
\gamma^*_{s}=\frac{2f(c\rightarrow c\bar{s})}{f(c\rightarrow c\bar{u}/
\bar{d})}(J=1)\approx\frac{2f(c\rightarrow D^{*+}_{s})}{f(c\rightarrow 
D^{*+})+f(c\rightarrow D^{*0})}
$$ 
and 
%-----------------------------------------------------------------------
$$
P^d_{V} =\frac{f(c\rightarrow c\bar{u}/\bar{d})(J=1)}{f(c\rightarrow c
\bar{u}/\bar{d})(J=0)}\approx\frac{f(c\rightarrow D^{*+})+f(c\rightarrow
 D^{*0})}{f(c\rightarrow D^{+})+f(c\rightarrow D^{0})}
$$
%-----------------------------------------------------------------------
were calculated from the fit results with the full error propagation and
taking into account the correlation between parameters.
\FB
%%%%%%%%%%%%%%%%%%%%%%%%%%%%%%%%%%%%%%%%%%%%%%%%%%%%%%%%%%%%%%%%%%%%%%%%
\section{Charm-quark fragmentation into hadrons in {\pmb \epem} 
collisions}
Measurements of charm-hadron-production cross-sections in $e^+e^-$ 
collisions hadrons were based on the differential momentum spectrum 
$\mathrm{d}\sigma(e^+e^- \rightarrow H)/\mathrm{d}x_p.$
The extrapolation to
the total cross-section was made in the original papers
using a theoretical
fragmentation function (e.g.\ Bowler~\cite{Bowler:1981sb} or 
Peterson~\cite{Peterson:1982ak}) 
\footnote{A proper extrapolation procedure requires 
only the  hadrons produced directly in fragmentation to be used in the 
fits. 
The hadrons produced in decays of excited charm hadrons should be 
treated separately.

In many cases the limited precision of the measurements 
makes this requirement hard to follow and the decay part of the meson 
production is treated together with the fragmentation part. 
In the cases 
where the contribution of hadrons from decays is comparable with the 
contribution of the direct production in fragmentation, e.g.\ for $D^0$ 
and $D^+$, the joint treatment could bias the results.
}. 

As mentioned before, the precise predictions of the total 
charm-production cross-section in $e^+e^-$ allow calculation of
 the fragmentation
 fractions without constraints on the sum of fractions.
This way the used hypothesis about the sum of fragmentation fractions 
(\eq{ffsumisone}) can be verified.
%%%%%%%%%%%%%%%%%%%%%%%%%%%%%%%%%%%%%%%%%%%%%%%%%%%%%%%%%%%%%%%%%%%%%%%%
\subsection{Charm-quark fragmentation fractions from measurements at 
${\pmb B}$-factories}
\label{sec:ssUPSILON}
The $B$-factories provided many results on   charm-hadron production 
around the $\Upsilon$ resonances, which can be used for the calculation 
of the charm-quark fragmentation fractions in hadrons (see 
\tab{EEUmeas}). 
The results  of the CLEO~\cite{Bortoletto:1988kw,Avery:1990bc}
 and ARGUS~\cite{Albrecht:1991ss,
Albrecht:1991pa,
Albrecht:1988an} 
experiments are represented as a product of the
charm-hadron cross-sections times decay branching ratios, 
$\sigma(e^+e^-\rightarrow H)\cdot {\cal B}(H\rightarrow 
\text{daughters}).$
The BELLE experiment~\cite{Seuster:2005tr} provided measurements of 
$\sigma(e^+e^-$\epjcbreak{}$\rightarrow  H)$. 
The BABAR experiment~\cite{Aubert:2006cp} provided a measurement of an 
average number of $\Lambda_c^+\rightarrow p K^-\pi^+$ decays per hadronic 
event 
\begin{multline*}
N^{q\bar{q}}_{\Lambda_c}\cdot{\cal B}(\Lambda_c^+\rightarrow p K^-\pi^+)
=2\frac{\sigma(e^+e^-\rightarrow  \Lambda_c^+)}{\sigma(e^+e^-\rightarrow 
\text{hadrons})} \times \epjcbreak{\times} {\cal B}(\Lambda_c^+
\rightarrow p K^-\pi^+)=\draftbreak{=}\arxivbreak{=}2R_c \cdot f(c\rightarrow
\Lambda_c^+) \cdot {\cal B}(\Lambda_c^+\rightarrow p K^-\pi^+),
\end{multline*}
where 
$R_{c}=\frac{\sigma(e^+e^-\rightarrow c\bar{c})}{\sigma(e^+e^-
\rightarrow \text{hadrons})}$ is the average number of charm-quark pairs
 per hadronic event.
A prediction of $R_c$ is needed to use this measurement as an input, 
therefore, as discussed in \sect{comb:proc}, it is  used only in the 
case of $\sigma(e^+e^- \rightarrow c\bar{c})$ fixed to a theoretical 
prediction.
\tabEEUmeas

For the calculation of the charm-quark fragmentation fractions a fit 
procedure is used as  described in \sect{comb}. The total 
charm-quark-production  cross-section is calculated as described in
 \app{appendixA}. 
The fit parameters 
are the fragmentation fractions. The obtained results are given  in the 
middle column of
\tab{EEUaverage}.
The sum of the charm-quark  fragmentation fractions into weakly decaying 
states calculated according to \eq{ffsumisone}, 
$S_{\Upsilon}=\SUpsilon,$ is consistent with unity.

\begin{table}[htbp]\tabfontsize\centering
\begin{tabular}{|c|c|c|}\hline
&Fixed $\sigma(e^+e^- \rightarrow c)$& Constrained $S$\\\hline\hline
$f(c\rightarrow{}D^{*+})$&$0.2470\pm0.0137$&$0.2525\pm0.0155$\\\hline
$f(c\rightarrow{}D^{*0})$&$0.2241\pm0.0304$&$0.2291\pm0.0316$\\\hline
$f(c\rightarrow{}D^{*+}_{s})$&$0.0532\pm0.0082$&$0.0544\pm0.0085$\\\hline
$f(c\rightarrow{}D^+)$&$0.2639\pm0.0139$&$0.2698\pm0.0125$\\\hline
$f(c\rightarrow{}D^0)$&$0.5772\pm0.0241$&$0.5901\pm0.0140$\\\hline
$f(c\rightarrow{}D^{+}_{s})$&$0.0691\pm0.0045$&$0.0707\pm0.0048$\\\hline
$f(c\rightarrow{}\Lambda_c^{+})$&$0.0526\pm0.0031$&$0.0611\pm0.0060$\\\hline\hline
$\chi^2{}$&$19.2$&$17.0$\\\hline
$n_{\text{dof}}{}$&$  21$&$  20$\\\hline\hline
$S$&$0.9701\pm0.0284$&$1.0000\pm0.0005$\\\hline
$R_{u/d}$&$0.9508\pm0.0752$&$0.9508\pm0.0752$\\\hline
$P^d_{V}$&$0.5601\pm0.0432$&$0.5601\pm0.0431$\\\hline
$\gamma_{s}$&$0.1644\pm0.0121$&$0.1644\pm0.0121$\\\hline
$\gamma^{*}_{s}$&$0.2257\pm0.0385$&$0.2257\pm0.0385$\\\hline

\end{tabular}
\caption{Average of charm-quark fragmentation fractions in hadrons in 
$e^+e^-$ collisions around $\sqrt{s}=\sqrtS\GeV$. The quantities $S$, $R_{u/d}$, $P^{d}_{V}$ and $\gamma_{s}$ were 
recalculated from the fit results taking into account correlation of fit parameters.
The value of minimised $\chi^2$  and the number degrees of freedom of the fit $n_{\text{dof}}$ are given as well.
}
\label{tab:EEUaverage}
\end{table}

The combination is also done according to \eq{ffwithsumh} and  imposing 
the constraint $S_{\Upsilon}-1=0$, to be consistent with the definition 
used for $e^{\pm}p$ and $pp$ data. The fit parameters are the 
fragmentation fractions and the total charm cross-section. 
The centre-of-mass energy dependence of the charm-quark cross-section 
is accounted for, according to formulae in \app{appendixA} taking the 
total charm-quark 
cross-section  at a centre-of-mass energy
 $\sqrt{s}=\sqrtS\GeV$  as a reference.
 The results are given  in \tab{EEUaverage}~(right column).
In this approach, the precise BABAR measurement of $\Lambda_c^+$ 
production~\cite{Aubert:2006cp} is not included in the combination
 since it requires usage of the $R_c$ theoretical prediction.
The latter has an  influence on other fragmentation-fraction results.
\FB
%%%%%%%%%%%%%%%%%%%%%%%%%%%%%%%%%%%%%%%%%%%%%%%%%%%%%%%%%%%%%%%%%%%%%%%%
\subsection{Charm-quark fragmentation fractions from measurements at 
LEP}
\label{sec:ssZZERO}
The LEP collider provided many results on the  charm-hadron production. 
The most valuable for the studies of fragmentation are results obtained 
from $Z$ decays. 
Most of those results are represented in the form of fraction of charm 
events  multiplied by branching ratios
$\frac{\Gamma(Z\rightarrow c\bar{c})}
{\Gamma(Z\rightarrow \text{hadrons})}$
$f(c\rightarrow H)\cdot {\cal B}(H\rightarrow\text{daughters})$
(see \tab{EEZmeas}).
In addition,  ALEPH~\cite{Barate:1999bg}, DELPHI~\cite{Abreu:1999vx} and 
OPAL~\cite{Ackerstaff:1997ki} provided measurements of $f(c \rightarrow 
D^{*+})$ from 
\tabEEZmeas
the fits of fragmentation functions (see \tab{EEZmeas}).  

For the calculation of charm-quark fragmentation fractions, a fit 
procedure is used, as  described in \sect{comb}. 
The theoretically 
calculated value that is used, $\frac{\Gamma(Z \rightarrow 
c\bar{c})}{\Gamma(Z \rightarrow \text{hadrons})}
=0.17223\pm0.00001$~\cite{Freitas:2014hra}, is in agreement with
the experimental world average 
$0.1721\pm0.003$~\cite{PDG2014}.
The fit parameters are the fragmentation fractions.
The results are given in the middle column of \tab{EEZaverage}.
The sum of the charm-quark  fragmentation fractions into  weakly 
decaying 
states calculated according to \eq{ffsumisone}, $S_{Z}=\SZ,$  
differs from unity by $\SZsign$ standard deviations.
\FB

\begin{table}[htbp]\tabfontsize\centering
\begin{tabular}{|c|c|c|}\hline
&Fixed $\frac{\Gamma_{c\bar c}}{\Gamma_{\text{had}}}$& Constrained $S$\\\hline\hline
$f(c\rightarrow{}D^{*+})$&$0.2369\pm0.0064$&$0.2454\pm0.0071$\\\hline
$f(c\rightarrow{}D^{*+}_{s})$&$0.0545\pm0.0144$&$0.0547\pm0.0145$\\\hline
$f(c\rightarrow{}D^+)$&$0.2267\pm0.0100$&$0.2429\pm0.0102$\\\hline
$f(c\rightarrow{}D^0)$&$0.5470\pm0.0215$&$0.5894\pm0.0132$\\\hline
$f(c\rightarrow{}D^{+}_{s})$&$0.0925\pm0.0082$&$0.0996\pm0.0083$\\\hline
$f(c\rightarrow{}\Lambda_c^{+})$&$0.0555\pm0.0065$&$0.0600\pm0.0066$\\\hline\hline
$\chi^2{}$&$ 6.7$&$ 7.8$\\\hline
$n_{\text{dof}}{}$&$  13$&$  13$\\\hline\hline
$S$&$0.9292\pm0.0261$&$1.0000\pm0.0005$\\\hline
$R_{u/d}$&$0.9987\pm0.0627$&$1.0348\pm0.0580$\\\hline
$P^d_{V}$&$0.6119\pm0.0185$&$0.6000\pm0.0177$\\\hline
$\gamma_{s}$&$0.2390\pm0.0224$&$0.2394\pm0.0223$\\\hline

\end{tabular}
\caption{Average of charm-quark fragmentation fractions into hadrons in  $Z$ decays.
 The quantities $S$, $R_{u/d}$, $P^d_{V}$ and 
$\gamma_s$ are recalculated from the fit results taking into account correlation of fit parameters.
The value of minimised $\chi^2$  and the number degrees of freedom of the fit $n_{\text{dof}}$ are given as well.}
\label{tab:EEZaverage}
\end{table}

The combination is also done using \eq{ffwithsumh}, and imposing 
the constraint $S_{Z}-1=0$, to be consistent with the definition 
used for $e^{\pm}p$ and $pp$ data. The fit parameters are the 
fragmentation 
fractions and the $\frac{\Gamma(Z \rightarrow c\bar{c})}{\Gamma(Z 
\rightarrow \text{hadrons})}$ ratio. The results, given in 
\tab{EEZaverage} (right column), 
are in good agreement with Ref.~\cite{Gladilin:2014tba}.
\FB
%%%%%%%%%%%%%%%%%%%%%%%%%%%%%%%%%%%%%%%%%%%%%%%%%%%%%%%%%%%%%%%%%%%%%%%%
\section{Charm-quark fragmentation into hadrons in {\pmb\ep} collisions}
\label{sec:ffep}
The charm-hadron-production cross-sections at HERA were measured in a 
restricted fiducial phase space. 
The extraction of the charm-quark fragmentation fractions  requires a 
special treatment, as described 
in detail in \app{appendixB}.
The approach followed in this analysis is  similar to the one originally 
used by the ZEUS collaboration~\cite{Chekanov:2005mm}.
%%%%%%%%%%%%%%%%%%%%%%%%%%%%%%%%%%%%%%%%%%%%%%%%%%%%%%%%%%%%%%%%%%%%%%%%
\FB
\subsection{Charm-quark fragmentation fractions from measurements in 
DIS}
Charm-quark fragmentation fractions in DIS in $e^{\pm}p$ collisions are 
calculated from ZEUS and H1 measurements given in \tab{DISmeas}.
\FB
\tabDISmeas
For the calculation of charm-quark fragmentation fractions a fit 
procedure is used as it is described in \sect{comb}.
The free parameters in the fit are the charm fragmentation fractions
and  pairs  of variables\epjcbreak{} $\sigma(c)_{i}|_{i=1\dots 3}$ and 
$\kappa_{i}|_{i=1\dots 3}$ for each set of measurement. 
Here, $\sigma(c)_{i}$ is the total charm cross-section in \ep, while 
$\kappa_{i}$ is the kinematic factor for decays from higher states 
(see \app{appendixB}). 
The parameter $\kappa$ is fixed to one for the low-$p_T$ measurements 
in Ref.~\cite{Abramowicz:2010aa} since the whole $p_T$ kinematic space 
was covered. 
The present beauty contributions in Ref.~\cite{Aktas:2004ka} 
have small impact on the result and are neglected.
The sum of charm fragmentation fractions $S_{ep~\text{DIS}}$ is 
constrained to unity.
The results of the averaging procedure are given in \tab{DISaverage}.

\begin{table}[htbp]\tabfontsize\centering
\begin{tabular}{|c|c|}\hline
         &    Constrained $S$ \\\hline\hline
$f(c\rightarrow{}D^{*+})$        &$0.2372\pm0.0173$\\\hline
$f(c\rightarrow{}D^+)$        &$0.2170\pm0.0203$\\\hline
$f(c\rightarrow{}D^0)$        &$0.6272\pm0.0287$\\\hline
$f(c\rightarrow{}D^{+}_{s})$        &$0.0945\pm0.0124$\\\hline
$f(c\rightarrow{}\Lambda_c^{+})$        &$0.0540\pm0.0195$\\\hline\hline
$\chi^2{}$        &$ 1.7$\\\hline
$n_{\text{dof}}{}$        &$   3$\\\hline\hline
$S$        &$1.0000\pm0.0004$\\\hline
$R_{u/d}$        &$1.2361\pm0.1331$\\\hline
$P^d_{V}$        &$0.6282\pm0.0440$\\\hline
$\gamma_{s}$        &$0.2240\pm0.0320$\\\hline

\end{tabular}
\caption{Average of charm-quark fragmentation fractions in $e^{\pm}p$ collisions in DIS.
 The quantities $S$, $R_{u/d}$, $P^d_{V}$ and 
$\gamma_s$ are recalculated from the fit results taking into account correlation of fit parameters.
The value of minimised $\chi^2$  and the number degrees of freedom of the fit $n_{\text{dof}}$ are given as well.}
\label{tab:DISaverage}
\end{table}

The obtained fragmentation fractions are in agreement with those 
obtained in the original publications~\cite{Chekanov:2007ch,
Aktas:2004ka}. 
The uncertainties of the obtained results  are somewhat larger  because 
this analysis, contrary to those studies, relies only on DIS results, 
whereas the HERA DIS papers~\cite{Chekanov:2007ch,
Aktas:2004ka} used fragmentation fractions into 
$\Lambda_c^+$ measured at $e^+e^-$ colliders.
\FB
%%%%%%%%%%%%%%%%%%%%%%%%%%%%%%%%%%%%%%%%%%%%%%%%%%%%%%%%%%%%%%%%%%%%%%%%
\subsection{Charm-quark fragmentation fractions from measurements in 
PHP}
Charm-quark fragmentation fractions in PHP in $e^{\pm}p$ collisions
 were calculated from measurements of  ZEUS 
            collaboration and given in \tab{PHPmeas}.
\FB
\tabPHPmeas
For the update of the latest ZEUS measurement~\cite{Abramowicz:2013eja} 
to the  decay branching ratios from \tab{BIGPDG}, the measured 
fragmentation fractions are first transformed into total charm-hadron 
cross-sections according to the formulae in \app{appendixB} and only 
then used in the calculations. 

\begin{table}[htbp]\tabfontsize\centering
\begin{tabular}{|c|c|}\hline
        &    Constrained $S$ \\\hline\hline
$f(c\rightarrow{}D^{*+})$        &$0.2345\pm0.0081$\\\hline
$f(c\rightarrow{}D^+)$        &$0.2341\pm0.0093$\\\hline
$f(c\rightarrow{}D^0)$        &$0.5991\pm0.0126$\\\hline
$f(c\rightarrow{}D^{+}_{s})$        &$0.0901\pm0.0062$\\\hline
$f(c\rightarrow{}\Lambda_c^{+})$        &$0.0675\pm0.0106$\\\hline\hline
$\chi^2{}$        &$ 5.2$\\\hline
$n_{\text{dof}}{}$        &$   4$\\\hline\hline
$S$        &$1.0000\pm0.0005$\\\hline
$R_{u/d}$        &$1.1209\pm0.0545$\\\hline
$P^d_{V}$        &$0.5970\pm0.0181$\\\hline
$\gamma_{s}$        &$0.2164\pm0.0162$\\\hline

\end{tabular}
\caption{Average of charm-quark fragmentation fractions in hadrons in $e^{\pm}p$ collisions in photoproduction.
 The quantities $S$, $R_{u/d}$, $P^d_{V}$ and 
$\gamma_s$ are recalculated from the fit results taking into account correlation of fit parameters.
The value of minimised $\chi^2$  and the number degrees of freedom of the fit $n_{\text{dof}}$ are given as well.}
\label{tab:PHPaverage}
\end{table}

In this procedure, the kinematic factor 
for decays from higher states, $\kappa$, is set to $ 1 $, since  the total 
phase space is considered from the fragmentation fraction definition, 
and the $\sigma(c)$ value cancels out in the procedure.
For the calculation of charm-quark fragmentation fractions, a fit 
procedure is used as it is described in \sect{comb}.
The free parameters in the fit are the charm fragmentation fractions
and  pairs  of variables $\sigma(c)_{i}|_{i=1 \dots 2}$ and 
$\kappa_{i}|_{i=1 \dots 2}$ for each set of measurement. 
Here, $\sigma(c)_{i}$ is the total charm cross-section in \ep, while 
$\kappa_{i}$ is the kinematic factor for decays from higher states 
(see \app{appendixB}). 
The sum of charm fragmentation fractions $S_{ep~\text{PHP}}$ is 
constrained to unity.
The results of the averaging procedure are given in \tab{PHPaverage}.
The obtained fragmentation fractions are in agreement with those 
obtained in the original publications~\cite{Chekanov:2005mm,
Abramowicz:2013eja}.
%
%
%%%%%%%%%%%%%%%%%%%%%%%%%%%%%%%%%%%%%%%%%%%%%%%%%%%%%%%%%%%%%%%%%%%%%%%%
\FB
\section{Charm-quark fragmentation into hadrons in 
${\pmb {pp}}$ collisions}
\label{sec:ffpp}
The ALICE experiment measured fiducial cross-sections of 
$D_s^{+}$~\cite{Abelev:2012tca} and differential $p_T$ cross-sections of 
$D^0$, $D^{+}$ and $D^{*+}$ mesons~\cite{ALICE:2011aa,Abelev:2012vra} at 
$\sqrt{s}=2.76\TeV$ and  $\sqrt{s}=7\TeV$. With an integration of the 
differential cross-sections of $D^0$, $D^+$ and $D^{*+}$
from Ref.~\cite{ALICE:2011aa} and $D^0$ from Ref.~\cite{Abelev:2012vra}, 
a coherent set of measurements in the kinematic range $2< p_T< 12\GeV$, $|y|< 0.5$ 
has been constructed (see Tab.~\ref{tab:LHCBmeas}) for the $\sqrt{s}=2.76\TeV$ 
and  $\sqrt{s}=7\TeV$. 
The LHCb experiment provided  measurements of charm-hadron 
cross-sections at $\sqrt{s}=7\TeV$~\cite{Aaij:2013mga}
and at $\sqrt{s}=13\TeV$~\cite{Aaij:2015bpa}. 
The ATLAS experiment recently measured the production cross-sections 
of $D^{*+}$, $D^+$ and $D_s^+$ mesons at $\sqrt{s}=7\TeV$~\cite{Aad:2015zix} 
in the kinematic range $3.5 < p_T< 20\GeV$, $|\eta|< 2.1$. 

The measurements together with the correlation matrix for
LHCb $\sqrt{s}=7\TeV$  are given in \tab{LHCBmeas}.

\tabLHCBmeasplusthirteenplusalice

For the calculation of charm-quark fragmentation fractions a fit 
procedure is used as it is described in \sect{comb}.
The free parameters in the fit are: the charm fragmentation fractions, the 
fiducial cross-sections for LHCb, ALICE and ATLAS measurements in kinematic regions 
given in \tab{LHCBmeas} and corresponding $\kappa$ parameters.

The constraint $S_{pp}-1=0$ is imposed.
As the Refs.~\cite{ALICE:2011aa,Abelev:2012vra} 
do not provide detailed decomposition of the systematic 
uncertainties, for every bin all systematic uncertainties were conservatively 
assumed to be fully correlated.
For all of these measurements  we assume the statistical and systematic 
uncertainties uncorrelated  and luminosity uncertainties -- fully 
correlated within a set of measurements at a given value of $\sqrt{s}$.

A set of orthogonal fully correlated uncertainties was obtained from the 
 covariance matrix of the $\sqrt{s}=7\TeV$ LHCb measurements with an 
 eigenvector  decomposition. The obtained uncertainties  are later 
 treated in the same way as other correlated sources in the combination.
The Ref.~\cite{Aaij:2015bpa} does not contain the correlation matrix 
for the measurements at $\sqrt{s}=13\TeV$, therefore simplified 
correlations between measurements were calculated as follows. All of 
the measurements include  $3.9\%$ fully correlated uncertainty related 
to luminosity included in the Ref.~\cite{Aaij:2015bpa}  to the 
systematic uncertainty. 
The systematic uncertainties also include the 
uncertainties on the branching ratios, which were treated correlated
with other branching-ratio uncertainties. The remaining systematic 
uncertainty were treated as fully uncorrelated for different 
measurements with the same $p_T$ cuts and fully correlated for the same 
measurements with different $p_T$ cuts. The statistical uncertainties 
of $\sigma(D^0)_{p_T<8\GeV}$, $\sigma(D^+)_{p_T<8\GeV}$ were split in 
two parts, which correspond to $p_T<1\GeV$ and $p_T>1\GeV$ regions. The 
later were considered fully correlated to the statistical uncertainties 
of the $\sigma(D^0)_{1<p_T<8\GeV}$ and $\sigma(D^+)_{1<p_T<8\GeV}$ 
measurements.
%

%To extrapolate the ATLAS measurement of the  $D^{+}_{s}\rightarrow 
%\phi(K^+K^-)\pi^+$ from the $|M(K^+K^-)-M(\phi(1020))|$$<7\MeV$ 
%kinematic region to all $\phi(1020)$ decays, the parametrisation of the 
%$M(K^+K^-)$ line-shape from Ref.~\cite{delAmoSanchez:2010yp} with 
%the particle masses from Ref.~\cite{PDG2014} was used. It was found 
%that $\phikksevenmev$\% of all decays involving $\phi$ fall into this 
%mass window.  This precedure results in $\ATLASFACTOR$ factor for the 
%central value and experimental uncertainties and reduces the 
%branching-ratio uncertainty  from $\pm 10\mb$ to $\pm \ATLASBR \mb$.

For the LHCb and ATLAS measurements of the  $D^{+}_{s}\rightarrow  K^+K^-\pi^+$ 
in the limited mass windows $M(K^+K^-)$  the following approach was used.
The branching ratios  were obtained from the integrals over 
the $M(K^+K^-)$ line shape that was parametrised as in Ref.~\cite{delAmoSanchez:2010yp}
with the total $D^{+}_{s}\rightarrow K^+K^-\pi^+$ signal normalised to 
${\cal B}(D^{+}_{s}\rightarrow K^+K^-\pi^+)=5.45\pm0.17\%$~\cite{PDG2014}. 
The results are given in Tab.~\ref{tab:BIGPDG}.

\begin{table}[hbtp]\tabfontsize\centering
\begin{tabular}{|c|c|}\hline
         &   Constrained $S$ \\\hline\hline
$f(c\rightarrow{}D^{*+})$        &$0.2337\pm0.0175$\\\hline
$f(c\rightarrow{}D^+)$        &$0.2274\pm0.0128$\\\hline
$f(c\rightarrow{}D^0)$        &$0.6176\pm0.0160$\\\hline
$f(c\rightarrow{}D^{+}_{s})$        &$0.0824\pm0.0084$\\\hline
$f(c\rightarrow{}\Lambda_c^{+})$        &$0.0639\pm0.0122$\\\hline\hline
$\chi^2{}$        &$ 6.9$\\\hline
$n_{\text{dof}}{}$        &$   7$\\\hline\hline
$S$        &$1.0000\pm0.0005$\\\hline
$R_{u/d}$        &$1.1913\pm0.1012$\\\hline
$P^d_{V}$        &$0.6059\pm0.0306$\\\hline
$\gamma_{s}$        &$0.1951\pm0.0216$\\\hline

\end{tabular}
\caption{Average of charm-quark fragmentation fractions in $pp$ collisions.
 The quantities $S$, $R_{u/d}$, $P^d_{V}$ and 
$\gamma_s$ are recalculated from the fit results taking into account correlation of fit parameters.
The value of minimised $\chi^2$  and the number degrees of freedom of the fit $n_{\text{dof}}$ are given as well.}
\label{tab:LHCBaverage}
\end{table}

The results of the fit are reported in \tab{LHCBaverage}. In addition to 
 the values of the fragmentation fractions, the fit delivers the
inclusive charm-production cross sections in the corresponding fiducial 
regions, which have particular interest.
Therefore, the values of these cross-sections obtained in the global 
combination with better precision are discussed below.
%%%%%%%%%%%%%%%%%%%%%%%%%%%%%%%%%%%%%%%%%%%%%%%%%%%%%%%%%%%%%%%%%%%%%%%%

\FB
\section{Selection of measurements for the extraction of fragmentation 
fractions}

The selection of the measurements for the extraction of fragmentation 
fractions was done according a set of criteria explained below.

First, the selection is limited to the measurements obtained in the 
collisions of high energy particle beams as it assures an absence of 
possible matter effects and the charm quark production mechanism in 
these environments is well understood.
The measurements of charm-hadron production in 
proton--meson, proton--nucleon and nucleon--nucleon 
collisions~\cite{Alves:1996rz,Barlag:1990bv,
Barlag:1990hg,Abt:2007zg,Tlusty:2012ix,Ye:2014eia} were omitted as 
those provide results in very specific production environment and energy 
ranges  which cannot be easily compared to the results in other 
experiments.

The second criteria of the selection is the  precision of the measured 
quantities:
 the measurements in $e^+e^-$ collisions with $\sqrt{s}=12-90\GeV$
 from 
 MARK-II~\cite{Yelton:1982ix}, HRS~\cite{Ahlen:1983gz,Derrick:1984ba,
 Derrick:1985ip,Low:1986nz,Baringer:1988ue}, TPC~\cite{Aihara:1985pp},
TASSO~\cite{Althoff:1983rt,Braunschweig:1989kq}, JADE~\cite{
Bartel:1984ud,Bartel:1985be}, VENUS~\cite{Hinode:1993gj} and some other 
experiments  have very limited precision and are not used for the 
global 
combination. 

The third criterion of the selection is the availability of sufficient 
measurements in the given physical environment needed for the extraction 
procedure. 
Several results  on charm production in $e^{\pm}p$ collisions (e.g.\ 
Ref.~\cite{Chekanov:2008yd}) and $pp$ collisions (e.g.\  
Refs.~\cite{Tlusty:2012ix,Ye:2014eia,Acosta:2003ax}) do not contain 
enough simultaneous 
measurements of hadron production and, therefore, cannot be treated 
independently
and/or constrain the results of the combination.
%%%%%%%%%%%%%%%%%%%%%%%%%%%%%%%%%%%%%%%%%%%%%%%%%%%%%%%%%%%%%%%%%%%%%%%%
\FB
\section{The global combination} 
\label{sec:ffglob}
To check the consistency of the data from different production regimes 
and also to extract the charm-quark fragmentation fractions with  
high precision, all input measurements introduced in the previous 
sections
 are used together to produce a global combination.
As discussed in \sect{ssUPSILON}, the $\Lambda_c^+$ measurement by 
the BABAR  experiment~\cite{Aubert:2006cp} is not included while 
obtaining the combined result. 
The free parameters of the fit are the charm-quark fragmentation 
fractions and  pairs of variables $\sigma(c)_{i}|_{i=1\dots 5}$ and 
$\kappa_{i}|_{i=1\dots 5}$  for three DIS and two PHP sets of 
measurements, $\frac{\Gamma_{c\bar c}}{\Gamma_{\text{had}}}$, 
$\sigma(e^+e^-\rightarrow c)$ at 
$\sqrt{s}=\sqrtS\GeV$, 
and the fiducial charm-quark cross-section and $\kappa$ parameters 
in $pp$ collisions, 
 at $\sqrt{s}=7\TeV$ and $\sqrt{s}=13\TeV$, 
corresponding to the phase space of the measurements. 
The constraint on the sum of the cross-sections of the weakly decaying
charm states, $S$, is imposed in the combination, i.e.\ the  
prediction for the total charm cross-sections in $e^+e^-$ collisions
is not used, in order to minimise model dependence
of the averaging procedure. 
The result of averaging $e^+e^-$, $e^{\pm}p$ and $pp$ data, with the 
constraint $S=1$ is presented in
the middle column of
\tab{FINALaverage} and is shown in \fig{FINALaverageFF}. The 
correlations between the fitted parameters are given in 
\tab{FINALcorrelations}.
The input data are in very good agreement with  $\chi^2 / n_\text{dof} 
=\fitgoodness.$
The result of the combination has significantly reduced uncertainties 
compared to individual measurements.

\figFINALaverageFF

\begin{table}[htbp]\tabfontsize\centering
\begin{tabular}{|c|c|c|}\hline
&\multirow{2}{*}{ Constrained $S$}&Constrained $S$, \\
&                & fixed $\sigma(e^+e^- \rightarrow c)$, $\frac{\Gamma_{cc}}{\Gamma_{\text{had}}}$.                                                                                   \\\hline\hline
$f(c\rightarrow{}D^{*+})$&$0.2429\pm0.0049$&$0.2386\pm0.0046$\\\hline
$f(c\rightarrow{}D^{*0})$&$0.2306\pm0.0315$&$0.2250\pm0.0299$\\\hline
$f(c\rightarrow{}D^{*+}_{s})$&$0.0548\pm0.0074$&$0.0537\pm0.0072$\\\hline
$f(c\rightarrow{}D^+)$&$0.2404\pm0.0067$&$0.2439\pm0.0067$\\\hline
$f(c\rightarrow{}D^0)$&$0.6086\pm0.0076$&$0.6141\pm0.0073$\\\hline
$f(c\rightarrow{}D^{+}_{s})$&$0.0802\pm0.0040$&$0.0797\pm0.0040$\\\hline
$f(c\rightarrow{}\Lambda_c^{+})$&$0.0623\pm0.0041$&$0.0549\pm0.0026$\\\hline\hline
$\chi^2{}$&$65.6$&$87.1$\\\hline
$n_{\text{dof}}{}$&$  64$&$  67$\\\hline\hline
$S$&$1.0000\pm0.0005$&$1.0000\pm0.0004$\\\hline
$R_{u/d}$&$1.0971\pm0.0354$&$1.1164\pm0.0354$\\\hline
$P^d_{V}$&$0.5578\pm0.0375$&$0.5403\pm0.0355$\\\hline
$\gamma_{s}$&$0.1890\pm0.0103$&$0.1859\pm0.0101$\\\hline
$\gamma^{*}_{s}$&$0.2314\pm0.0347$&$0.2316\pm0.0346$\\\hline

\end{tabular}
\caption{Average of charm-quark fragmentation fractions in hadrons. The quantities $S$, $R_{u/d}$, $P^d_{V}$ and 
$\gamma_s$ are recalculated from the fit results taking into account correlation of fit parameters.
The value of minimised $\chi^2$  and the number degrees of freedom of the fit $n_{\text{dof}}$ are given as well.}
\label{tab:FINALaverage}
\end{table}

\begin{table}[htbp]\tabfontsize\centering
\begin{tabular}{|c|ccccccc|}\hline
%&$f(D^{*+})$&$f(D^{*0})$&$f(D^{*+}_{s})$&$f(D^+)$&$f(D^0)$&$f(D^{+}_{s})$&$f(\Lambda_c^{+})$\\\hline\hline
&$D^{*+}$&$D^{*0}$&$D^{*+}_{s}$&$D^+$&$D^0$&$D^{+}_{s}$&$\Lambda_c^{+}$\\\hline\hline
$D^{*+}$&1.00&-0.02&-0.02&-0.08&0.19&-0.07&-0.12\\
$D^{*0}$&-0.02&1.00&0.02&-0.07&0.07&0.01&-0.01\\
$D^{*+}_{s}$&-0.02&0.02&1.00&-0.05&-0.07&0.23&-0.01\\
$D^+$&-0.08&-0.07&-0.05&1.00&-0.66&-0.19&-0.19\\
$D^0$&0.19&0.07&-0.07&-0.66&1.00&-0.32&-0.41\\
$D^{+}_{s}$&-0.07&0.01&0.23&-0.19&-0.32&1.00&-0.07\\
$\Lambda_c^{+}$&-0.12&-0.01&-0.01&-0.19&-0.41&-0.07&1.00\\

\hline
\end{tabular}
\caption{Correlation of charm-quark fragmentation fractions from the fit with constrained $S$. }
\label{tab:FINALcorrelations}
\end{table}

As an alternative, the combination is also performed using both the 
constraint on $S$ as well as theoretical predictions of charm 
production
in $e^+e^-$ collisions and $Z$ decays, i.e.\  $\sigma(e^+e^-\rightarrow c)$
at $\sqrt{s}=\sqrtS\GeV$
and $\frac{\Gamma_{c\bar c}}{\Gamma_{\text{had}}}$. 
This approach also allows to include the precise BABAR 
measurement of $\Lambda_c^+$ production~\cite{Aubert:2006cp} 
using the $R_c$ calculation as described in \app{appendixA}, 
which significantly affects the averaged value of $f(\Lambda_c^+)$.
The result of the averaging procedure with this approach
 is given in the right column of \tab{FINALaverage} for completeness.
The result is more model dependent than the default 
combination, but has a higher precision. At the same time, the result 
visibly differs from the result of the default procedure.
This may partially be
traced to the value $S_{Z} = \SZ$ for the accurate LEP
measurements, which differs markedly from $1$ (see \sect{ssZZERO}).
This difference is also reflected in the larger $\chi^2 /n_{\text{dof}}$
value compared to the default global combination.
The difference in the $f(c \rightarrow \Lambda_c^+)$ precision is to
a large extent due to inclusion of the precise BABAR 
 data~\cite{Aubert:2006cp}.

\figFINALaverageRPg
The extracted  $R_{u/d}$, $P^d_{V}$ and $\gamma_{s}$ factors 
are provided in \tab{FINALaverage} and shown in 
\fig{FINALaverageRPg}.
The combined data are also compared to recent 
measurements~\cite{Acosta:2003ax, David:2007iv}
that were not included in the combination.
In particular, $R_{u/d} = \RudG\,$ is in fair agreement with the isospin 
invariance hypothesis $R_{u/d} = 1$ within $\RudGsign$ standard 
deviations.
The values of the $\sigma(pp\rightarrow c)$ cross-sections, obtained in 
the global fit (see \tab{CSLHCB})
are consistent with those obtained in the original analysis,
but have significantly reduced uncertainties.
\tabCSLHCBplusalice
The consistent treatment of the LHCb and ALICE measurements in the 
combination  procedure allows unbiased calculation of the ratio of the 
inclusive fiducial charm-quark production cross-sections:
\begin{multline*}
R_{7/2.76}=\frac{\sigma(pp\rightarrow c)_{7\TeV,2< p_T< 12\GeV,
|y|< 0.5}}{\sigma(pp\rightarrow c)_{2.76\TeV,2< p_T< 12\GeV,|y|< 
0.5}}=\epjcbreak{=}\Rseventwosevensix
\end{multline*}
and
$$R_{13/7}=\frac{\sigma(pp\rightarrow c)_{13\TeV,p_T< 
8\GeV,2<y<4.5}} {\sigma(pp\rightarrow c)_{7\TeV,p_T< 
8\GeV,2<y<4.5}}=\Rthirteenseven.$$
The $R_{7/2.76}$ is compatible with the predictions in 
Ref.~\cite{Averbeck:2011ga}.
The  $R_{13/7}$ value is visibly higher than the 
theoretical prediction
$R_{13/7}{\rm(theory)} = 1.39^{+0.12}_{-0.29}$~\cite{Gauld:2015yia}. 
\FB
%%%%%%%%%%%%%%%%%%%%%%%%%%%%%%%%%%%%%%%%%%%%%%%%%%%%%%%%%%%%%%%%%%%%%%%%
\section{Excited states}  
In addition to the average fragmentation fractions for the ground, 
$L=0$, states, some fragmentation fractions for the excited, 
$L=1$ charm hadrons are calculated.

The  measurements used for the calculations are shown in  \tab{excited}.
The unpublished measurement of $f(c\rightarrow D^+_{s1})$ from 
Ref.~\cite{Verbytskyi:2013jsa} was not used.
The fragmentation fractions were not updated to the most recent 
branching ratios, as the difference between the used branching ratios 
and the newest is  negligible in comparison to statistical and 
systematical  uncertainties of the measurements, and is well below the 
given numerical precision of the individual measurements in 
\tab{excited}.

\tabexcited
The averages are calculated with an assumption of fully uncorrelated 
statistical and systematical uncertainties. 

\begin{table}[htbp]\tabfontsize\centering
\begin{tabular}{|c|c|}\hline
        &    Average  $(10^{-2})$ \\\hline\hline
$f(c\rightarrow{}D^+_{1})$        &$4.60{}^{+2.69}_{-1.82}$\\\hline
$f(c\rightarrow{}D^{*+}_{2})$        &$3.20{}^{+0.94}_{-0.82}$\\\hline
$f(c\rightarrow{}D^0_{1})$        &$2.97\pm0.38$\\\hline
$f(c\rightarrow{}D^{*0}_{2})$        &$3.94\pm0.68$\\\hline
$f(c\rightarrow{}D^{+}_{s1})$        &$1.09\pm0.14$\\\hline\hline
$\gamma_{s1}$        &$28.7{}^{+ 8.0}_{-10.9}$\\\hline

\end{tabular}
\caption{Average of charm-quark fragmentation fractions in excited charm
 mesons. The $\gamma_{s1}$ quantity is calculated from the averaging results without taking into account correlations.}
\label{tab:FINALexcited}
\end{table}

The results of the averaging 
procedure are given in \tab{FINALexcited}.
The strangeness-suppression  factor 
for $L=1,J=1^+$ charm mesons is calculated neglecting $D(2430)^0$ 
contribution and  assuming  $D^+_1$ is $1^{+}$ state: 
$$
\gamma_{s1}\approx\frac{2f(c\rightarrow D^+_{s1})}
{f(c\rightarrow D^0_1)+f(c\rightarrow D^+_1)}.
$$ 
%%%%%%%%%%%%%%%%%%%%%%%%%%%%%%%%%%%%%%%%%%%%%%%%%%%%%%%%%%%%%%%%%%%%%%%%
\section{Summary}                      
\label{sec:con}
A summary of measurements of the fragmentation of charm quarks into a
specific charm hadron is given. 
The analysis includes data collected in photoproduction and deep 
inelastic scattering in $e^{\pm}p$ collisions and well as $e^+e^-$ and $pp$
 data.
Measurements in different production regimes agree within uncertainties, 
supporting the hypothesis that fragmentation proceeds independent of the 
specific production process. 
Averages of the fragmentation fractions are presented. 
The global average has significantly reduced uncertainties compared 
to individual measurements.
In addition, the hypothesis that the sum of fragmentation fractions 
of all known weakly decaying charm hadrons is equal to unity is
checked to hold within $3$ standard deviations 
using the $e^+e^-$ data.
%
%%%%%%%%%%%%%%%%%%%%%%%%%%%%%%%%%%%%%%%%%%%%%%%%%%%%%%%%%%%%%%%%%%%%%%%%
\section*{Acknowledgements}
\label{sec:ack}
We thank Erich Lohrmann for his major contribution to the development 
of this paper. We thank  Uri Karshon and Stefan Kluth  for useful 
discussions and help in the work with the bibliography.
We also thank  Alexander  Glazov  and Ian Brock
for the critical reading of the manuscript and 
useful suggestions on text improvement.
\FB
%\newpage
\begin{appendices}
%%%%%%%%%%%%%%%%%%%%%%%%%%%%%%%%%%%%%%%%%%%%%%%%%%%%%%%%%%%%%%%%%%%%%%%%
\section{Predictions for charm production at ${\pmb B}$-factories}
\label{sec:appendixA}
The  total   cross-section of quark $q$ production in $e^+e^-$ collisions 
at energies $2M(q) \ll \sqrt{s} \ll M(Z)$ can be given as
\begin{equation}\label{eq:cseeqq}
\sigma(e^+e^-\rightarrow q)=
2\sigma(e^+e^-\rightarrow l^+l^-)\sum_{\text{colours}} v_q^2  r_q,
\end{equation}
where $v_q$ is the vector electromagnetic coupling of the quark $q$ 
(i.e.\  charge), $r_q(s)$ are the correction coefficients with higher 
order QCD corrections and
\begin{equation}\label{eq:cs}
\sigma(e^+e^-\rightarrow l^+l^-)=4\alpha^2(s)\pi/3s
\end{equation}
is the  total  cross-section of massless charged lepton pair production.
In this work, the calculations of the $r_q(s)$ were done according to 
Ref.~\cite{Chetyrkin:2000zk} at the reference energy of 
$\sqrt{s}=\sqrtS\GeV$  and assuming the $c$ quark is the heavy one. The 
constants used for the calculations in \eq{cseeqq} and \eq{cs} are
the strong coupling $\alpha_s(\sqrt{s}=\sqrtS\GeV)=0.172$
~\cite{Chetyrkin:2000zk}, the $\overline{\text{MS}}$ charm-quark mass 
$m_c(\sqrt{s}=\sqrtS\GeV)=0.74\GeV$~\cite{Chetyrkin:2000zk} and 
the electromagnetic coupling $\alpha(\sqrt{s}=\sqrtS\GeV)=1/132.0$ 
(calculated according to~\cite{Altarelli:1989hx,Burkhardt:1989ky} as 
implemented in~\cite{Harris:1997zq}). The uncertainties on the given 
values are negligible. The result of the calculations is 
$\sigma(e^+e^-\rightarrow c,\sqrt{s}=\sqrtS\GeV)=\sigmaeecc\nb$.
To verify the calculations,  
\eq{cseeqq}  can be rewritten as 
\begin{equation*}\label{eq:sigmactwo}
\sigma(e^+e^-\rightarrow c)=2  
\sigma(e^+e^-\rightarrow l^{+}l^{-}) R_c R_{\text{had}},
\end{equation*} where the quantities
\begin{equation*}\label{eq:rh}
R_{\text{had}}=
\frac{\sigma(e^+e^-\rightarrow \text{hadrons})}{\sigma(e^+e^-\rightarrow 
l^+l^-)}=\sum\limits_{u,d,s,c}\sum\limits_{\text{colours}} v_q^2 
r_q=\Rhadrons
\end{equation*}
and 
\begin{equation*}\label{eq:rc}
R_{c}=\frac{\sigma(e^+e^-\rightarrow c\bar{c})}{\sigma(e^+e^-\rightarrow 
\text{hadrons})}=\frac{\sum\limits_{\rm colours} v_c^2 r_c}
{\sum\limits_{u,d,s,c}\sum\limits_{\rm colours} v_q^2 r_q}=\Rc
\end{equation*}
can be compared with the existing measurements and predictions.
It was found that $R_{\text{had}}$ is in agreement with the direct  
measurement from CLEO below $\sqrt{s}=10.56\GeV$
$R_{\text{had,CLEO}}=3.591\pm0.003\pm0.067\pm0.049$
~\cite{Besson:2007aa} 
and $R_{c}$ is in  agreement with the CLEO Monte-Carlo based estimation
$R_{\text{c,CLEO}}=0.37\pm0.05$~\cite{Bortoletto:1988kw}.
For all the theoretically calculated values, the 
uncertainties of calculations are negligible.
\FB
%%%%%%%%%%%%%%%%%%%%%%%%%%%%%%%%%%%%%%%%%%%%%%%%%%%%%%%%%%%%%%%%%%%%%%%%
\section{Extraction of the charm-quark fragmentation fractions from the measurements 
in the restricted phase space}
\label{sec:appendixB}
Often the production cross-sections of charm hadrons 
are measured in some restricted (e.g.\ in transverse 
momentum and  pseudorapidity) kinematical region $v$ and cannot be 
used directly (i.e.\ without extrapolation to the full kinematical space)
 in the \eq{ffwithsumh}. To avoid the extrapolation and obtain 
 unbiased charm-quark fragmentation fractions, the following approach is
  used.

The full   cross-section of a  charm hadron, $\sigma(H)_{\in v}$ can be
 split into  cross-section of direct 
production $\sigma(H)_{\text{dir}, \in v}$ and the 
contribution from the decays of heavier charm states  $H^*$,
$\sigma( H)_{\text{decays},  \in v}$:
\begin{multline*}
\sigma(H)_{{\in v}}= 
\sigma(H)_{\text{dir},{\in v}}+
\sigma(H)_{\text{decays},{\in v}}=\epjcbreak{=}
 \sigma(H)_{\text{dir},{\in v}}+
\sum_{all\, H^*}\sigma(  H^{*})    
{\cal B}(H^{*}\rightarrow H)k_{H^*\rightarrow H},
\end{multline*}
where $k_{H^*\rightarrow H}<1$ is a fraction of $H^{*} \rightarrow H$ 
decays with $H$  in $v$. 
The lack of experimental data 
in this analysis  allows to  consider only the  heavier charm states 
that are giving the largest contribution to the total 
cross-sections. For this reason all $D^0$ and $D^+$   produced not in 
$D^{*+}$ and $D^{*0}$ decays are considered as produced directly. 
Because of similar kinematics of $D^{*}\rightarrow D$ decays it is also 
assumed that $k_{D^*\rightarrow D}=k$. The measurements on heavier 
charm baryons  and charm strange mesons are absent, 
but all of those decay dominantly to 
$\Lambda_c^+$ and $D^{+}_s$ so we treat all 
$\Lambda_c^+$ and  $D^{+}_s$ as produced directly. 
Within a common phase space, the impact of the different masses 
of the hadrons on the fragmentation process is neglected.

With the assumptions above, we have the
equations: 
$$
\left\lbrace
\def\arraystretch{1.15}
\array{lcl}
\sigma( D^+)_{\in v}        &=&
\sigma(  D^+)_{\text{dir},\in v}+k\sigma(D^{*+})_{\text{dir}}{\cal B}(D^{*+} \rightarrow D^+)\\
\sigma(D^0)_{\in v}         &=&\sigma(D^0)_{\text{dir},D^0\in v}+k \sigma(  D^{*0})_{\text{dir}}{\cal B}(D^{*0} \rightarrow D^0)\\
                            &&+k\sigma(  D^{*+})_{\text{dir}}{\cal B}(D^{*+}\rightarrow D^0)\\
\sigma( D^+_s)_{\in v}      &=&\sigma(  D^+_s)_{\text{dir},\in v}\\
\sigma( \Lambda_c^+)_{\in v}&=&\sigma(  \Lambda_c^+)_{\text{dir},\in v}\\
\sigma( D^{*+})_{\in v}     &=&\sigma(  D^{*+})_{\text{dir},\in v}\\
\sigma( D^{*0})_{\in v}     &=&\sigma(  D^{*0})_{\text{dir},\in v}.\\
\endarray\right.
$$
Assuming  $\sigma(H)_{\text{dir},\in v}=\sigma( c)_{\in v}
f(c \rightarrow H)_{\text{dir}}$ and introducing 
$\kappa=k\sigma( c)/\sigma( c)_{\in v}$
 we have:	
$$
\left\lbrace
\def\arraystretch{1.15}
\array{lcl}
\sigma( D^+)_{\in v}        &=&\sigma( c)_{\in v}(f(c \rightarrow  D^+)_{\text{dir}}\\
                            &&+\kappa f(c \rightarrow   D^{*+})_{\text{dir}}{\cal B}(D^{*+} \rightarrow D^+))\\
\sigma( D^0)_{\in v}        &=&\sigma( c)_{\in v}(f(c \rightarrow   D^0)_{\text{dir}}\\
                            &&+\kappa f(c \rightarrow   D^{*0})_{\text{dir}}{\cal B}(D^{*0} \rightarrow D^0)\\
                            &&+\kappa f(c \rightarrow D^{*+})_{\text{dir}}{\cal B}(D^{*+} \rightarrow D^0))\\
\sigma( D^+_s)_{\in v}      &=&\sigma( c)_{\in v}f(c \rightarrow   D^+_s)_{\text{dir}}\\
\sigma( \Lambda_c^+)_{\in v}&=&\sigma( c)_{\in v}f(c \rightarrow   \Lambda_c^+)_{\text{dir}}\\
\sigma( D^{*+})_{\in v}     &=&\sigma( c)_{\in v}f(c \rightarrow   D^{*+})_{\text{dir}}\\
\sigma( D^{*0})_{\in v}     &=&\sigma(c)_{\in v}f(c \rightarrow   D^{*0})_{\text{dir}}.
\endarray\right.
$$
In the full kinematical space:
$$
\left\lbrace
\def\arraystretch{1.15}
\array{lcl}
f(c \rightarrow  D^+)   &=&f(c \rightarrow  D^+)_{\text{dir}}\\
                        &&+f(c \rightarrow   D^{*+})_{\text{dir}}{\cal B}(D^{*+} \rightarrow D^+)\\
f(c \rightarrow  D^0)   &=&f(c \rightarrow   D^0)_{\text{dir}}\\
                        &&+f(c \rightarrow   D^{*0})_{\text{dir}} {\cal B}(D^{*0} \rightarrow D^0)\\
                        &&+f(c \rightarrow   D^{*+})_{\text{dir}}{\cal B}(D^{*+} \rightarrow D^0)\\
f(c \rightarrow  D^+_s) &=&f(c \rightarrow   D^+_s)_{\text{dir}}\\
f(c \rightarrow  \Lambda_c^+) &=&f(c \rightarrow   \Lambda_c^+)_{\text{dir}}\\
f(c \rightarrow  D^{*+})      &=&f(c \rightarrow   D^{*+})_{\text{dir}}\\
f(c \rightarrow  D^{*0})      &=&f(c \rightarrow   D^{*0})_{\text{dir}}.
\endarray\right.
$$
In general, to solve the system the measurements of $D^{*0}$ production 
are needed. However, these can be avoided with an assumption of isospin 
invariance:$$\frac{f(c\rightarrow D^+)_{\text{dir}}}
{f(c\rightarrow D^0)_{\text{dir}}}
=\frac{f(c\rightarrow D^{*+})_{\text{dir}}}{f(c\rightarrow 
D^{*0})_{\text{dir}}}.$$
The last two systems are the working equations for the calculation of 
the  charm fragmentation fractions from the cross-section measurements 
in the restricted phase space.
\end{appendices}

{\bibliographystyle{./FFC}{\raggedright\bibliography{FFC.bib}}}\vfill\eject
\clearpage
\end{document}